\documentclass[reprint,floatfix, twocolumn,superscriptaddress, longbibliography]{revtex4-1}
\usepackage{latexsym}
\usepackage{lineno}
\usepackage{hyperref}
\usepackage{url}
\usepackage[caption=false]{subfig}
\usepackage{graphicx}
\usepackage{verbatim}
\usepackage{multirow}
\usepackage{amsmath, bm}
\newcommand{\beq}{\begin{eqnarray}}
\newcommand{\eeq}{\end{eqnarray}}
\usepackage{mathrsfs}
\usepackage{float}
\usepackage{dsfont}
\usepackage{bbm}
\usepackage[usenames, dvipsnames]{color}
\usepackage{mathtools}
\usepackage{slashed}
\usepackage{physics}	
\usepackage{graphicx}   
\usepackage{epstopdf}
\usepackage{hyperref}   
\usepackage{bbold}
\usepackage{wasysym}
\usepackage{amssymb}
\usepackage{feynmp}
\usepackage[symbol]{footmisc}
\usepackage[latin1]{inputenc}
\usepackage{tikz}
\usetikzlibrary{decorations.pathmorphing}
\usetikzlibrary{shapes.misc}
\tikzset{cross/.style={cross out, draw=black, minimum size=8*(#1-\pgflinewidth), inner sep=0pt, outer sep=0pt},
cross/.default={1pt}}
\usetikzlibrary{patterns,math}

\usepackage[normalem]{ulem}

\newcommand{\RN}[1]{%
  \textup{\uppercase\expandafter{\romannumeral#1}}%
}

\newcommand{\sur}[1]{{\color{black}  #1}}

\newcommand{\ncmd}{\newcommand}
\ncmd{\nn}{\nonumber}
\ncmd{\mbf}[1]{\bs{#1}}
\ncmd{\gam}{\gamma}
\ncmd{\sig}{\sigma}
\ncmd{\pha}{\alpha}
\ncmd{\lam}{\lambda}
\ncmd{\dl}{\delta}
\ncmd{\kap}{\kappa}
\ncmd{\Lam}{\Lambda}
\ncmd{\Gam}{\Gamma}
\ncmd{\Dl}{\Delta}
\ncmd{\Ups}{\Upsilon}
\ncmd{\Om}{\Omega}
\ncmd{\eps}{\epsilon}
\ncmd{\veps}{\varepsilon}
\ncmd{\vphi}{\varphi}
\ncmd{\vtheta}{\vartheta}
\ncmd{\tw}{\text{w}}
\ncmd{\pll}{\parallel}
\ncmd{\mc}{\mathcal}
\ncmd{\mf}{\mathfrak}
\ncmd{\bs}{\boldsymbol}
\usepackage{pifont}
\usepackage[T1]{fontenc}
\usepackage{bclogo}
\ncmd{\note}[1]{{\color{red}{\ding{168} #1}}}
\ncmd{\eq}[1]{Eq. \eqref{#1}}
\ncmd{\fig}[1]{Fig. \ref{#1}}
\ncmd{\suppl}{\note{`Supplementary Information'}}
\ncmd{\pg}[1]{\textcolor{red}{#1}}

\newcommand{\vk}{{\bm{k}}}

\newcommand{\vq}{{\bm{q}}}

\begin{document}
\title{
Electronic properties, correlated topology and Green's function zeros
}
\begin{abstract}
There is extensive current interest
about
electronic topology in correlated settings.
In strongly correlated systems, 
contours of Green's function zeros may develop in frequency-momentum space, and their role in correlated topology has increasingly been recognized.
However, whether and how 
the zeros 
contribute to electronic properties is a matter of uncertainty.
Here we 
address the issue in 
an exactly solvable model for Mott insulator.
We show that 
the
Green's function zeros contribute to
several physically measurable correlation functions, 
in a way that does not run into inconsistencies. In particular, the physical properties remain
robust to chemical potential variations up to the Mott gap
as it should be based on general considerations. 
Our work sets the stage for further understandings on the rich interplay among topology, symmetry and strong correlations.
\end{abstract}

\author{Chandan Setty$^{\dagger}$}
\thanks{These two authors contributed equally.}
\affiliation{Department of Physics and Astronomy, Rice Center for Quantum Materials, Rice University, Houston, Texas 77005, USA}
\affiliation{Department of Physics and Astronomy, Iowa State University, Ames, Iowa 50011, USA}
\affiliation{Ames National Laboratory, U.S. Department of Energy, Ames, Iowa 50011, USA}

\author{Fang Xie$^{\oplus}$}
\thanks{These two authors contributed equally.}
\affiliation{Department of Physics and Astronomy, Rice Center for Quantum Materials, Rice University, Houston, Texas 77005, USA}

\author{Shouvik Sur}
\affiliation{Department of Physics and Astronomy, Rice Center for Quantum Materials, Rice University, Houston, Texas 77005, USA}

\author{Lei Chen}
\affiliation{Department of Physics and Astronomy, Rice Center for Quantum Materials, Rice University, Houston, Texas 77005, USA}
 
\author{Maia\ G.\ Vergniory}
\affiliation{Donostia International  Physics  Center,  P. Manuel  de Lardizabal 4,  20018 Donostia-San Sebastian,  Spain}
\affiliation{Max Planck Institute for Chemical Physics of Solids, Noethnitzer Str. 40, 01187 Dresden, Germany}

\author{Qimiao Si}
\affiliation{Department of Physics and Astronomy, Rice Center for Quantum Materials, Rice University, Houston, Texas 77005, USA}

\maketitle

\section{Introduction}
In noninteracting systems, electronic topology is formulated within band theory. Recent years have seen systematic development on how symmetries
of crystalline lattices constrain 
 topology and how they can be utilized to search for new topological materials~\cite{Armitage2017,Nagaosa2020, Bradlyn2017,Cano2018,Po2017,Watanabe2017,cano2021band}.
In interacting settings, symmetry constraints have been considered in terms of Green's functions, either through a renormalized particle picture~\cite{Iraola21, Lessnich2021,Soldini2022}
in the form of a topological Hamiltonian~\cite{
Wang-Zhang_PRX2012, Wang-Yan2013} 
or by recognizing that the eigenvectors of the  
exact Green's function 
in a many-body system
form a representation of lattice space group~\cite{Hu-Si2021}.
The latter approach, 
which was introduced in the context of Weyl-Kondo semimetal~\cite{Lai2018,Grefe2020,
Dzsaber2017,Dzs-giant21.1}
 and provided the theoretical basis for its robustness~\cite{Chen-Si2022}, 
has led to the realization~\cite{Setty2023} that Green's function zeros~\cite{Dzyaloshinskii2003} of an interacting lattice system obey symmetry constraints;
accordingly, the Green's function zeros participate in the formation of
correlated electronic topology, just as Green's function poles do.
Concurrently, 
the role of
Green's function zeros has been studied 
in the context of the edge spectrum of 
interacting topological insulators~\cite{wagner2023mott}.

Quasiparticles represent the low energy excitations of a Fermi liquid. The quasiparticles are conveniently described in terms of a Green's function approach, in which they appear as
poles of the single particle 
Green's function~\cite{AGD}.
They are characterized by the quasiparticle weight -- a quantity that plays a central role in 
the microscopic Fermi liquid
theory. 
When electron correlations are strong, the quasiparticle weight 
becomes very small,
and its 
effect on observables is well documented;
exemplary settings can be found in Refs.~\cite{Kotliar2006-RMP, 
Florens-Georges2004, 
Yu-Si2012,Paschen-Si-2021}.
In the extreme correlation limit, the quasiparticle weight may vanish leading to 
a
breakdown of the Fermi liquid. 
The precise manner in which thermodynamic and transport properties are affected 
by interactions
in this limit  
is a central question 
in the field of correlation physics.
\par
Mott insulators (MIs) occupy a special place in the physics of strongly correlated systems
~\cite{Fazekas1999}. In MIs, the quasiparticle weight as well as the single particle Green's function vanish 
to yield Green's function zeros along certain frequency-momentum contours. 
There
has been considerable debate regarding the role of 
zeros on the physical charge and correlation functions~\cite{Dzyaloshinskii2003, AGD, Volovik2003,Rosch2007, Gurarie2011, Gurarie2011-2, Phillips2013, Yunoki2017, Setty2020, Setty2021, Setty2021-Kondo, Xu2021, Fabrizio2022, Fabrizio2023,Phillips2023}. 
\par
Like poles
across a Fermi surface, the real part of the Green's function changes sign across 
a zero 
surface;
hence they
contribute to the Luttinger volume~\cite{Dzyaloshinskii2003, AGD} and single particle winding numbers~\cite{Yunoki2017}. Similarly, zeros are key to the generalization of index theorems~\cite{Volovik2003} to interacting settings, and play an essential role in 
understanding symmetric mass generation~\cite{Vishwanath2018, Xu2021, You2023-1, You2023-2}. 
In interacting topological insulators, it was argued that zeros allow for topological transitions to occur without closing the boundary gap~\cite{Gurarie2011, Gurarie2011-2, Xu2015}.
\par 
While these properties raise the prospect of 
the
zeros being experimentally measurable, their relationship to observables has been tenuous at best.
For example, it is required that physical properties are independent of chemical potential variations up to the Mott scale at zero temperature~\cite{Rosch2007} despite zeros occurring in the insulating gap. 
In addition, when determining the zeros' contribution to physically measurable correlations, it is crucial to 
keep track of conservation laws and the associated Ward identities.
It is important to address these issues 
in order to properly assess the contributions of Green's function zeros to physical properties.
Furthermore, Green's function zeros are also difficult to probe experimentally. By definition, the vanishing spectral weight makes their detection challenging. 
Sharpening the theoretical understanding about how the Green's function zeros affect physical correlation functions is expected to be important for probing the zeros experimentally in the future.
\par 
In this work, we argue that certain robust physical properties of 
electron systems 
can indeed 
capture the contributions of
the Green's function zeros 
and, importantly, they do
so in a way that is
consistent with the aforementioned expectations. These properties can be exploited to indirectly gather properties that may otherwise be elusive to conventional probes.
Our focus here is to illustrate how zeros \textit{contribute} to observables in a consistent, gauge invariant manner, rather than propose a concrete experimental set-up where they can be systematically extracted. 
We demonstrate our claims by considering an exactly solvable model of a Mott insulator (MI)~\cite{HK1992} as a prototypical example where extreme correlation effects are realized~\cite{Fazekas1999}. \par
\begin{figure}[h!]
     \centering
     \includegraphics[width=8cm, height=6cm, keepaspectratio]{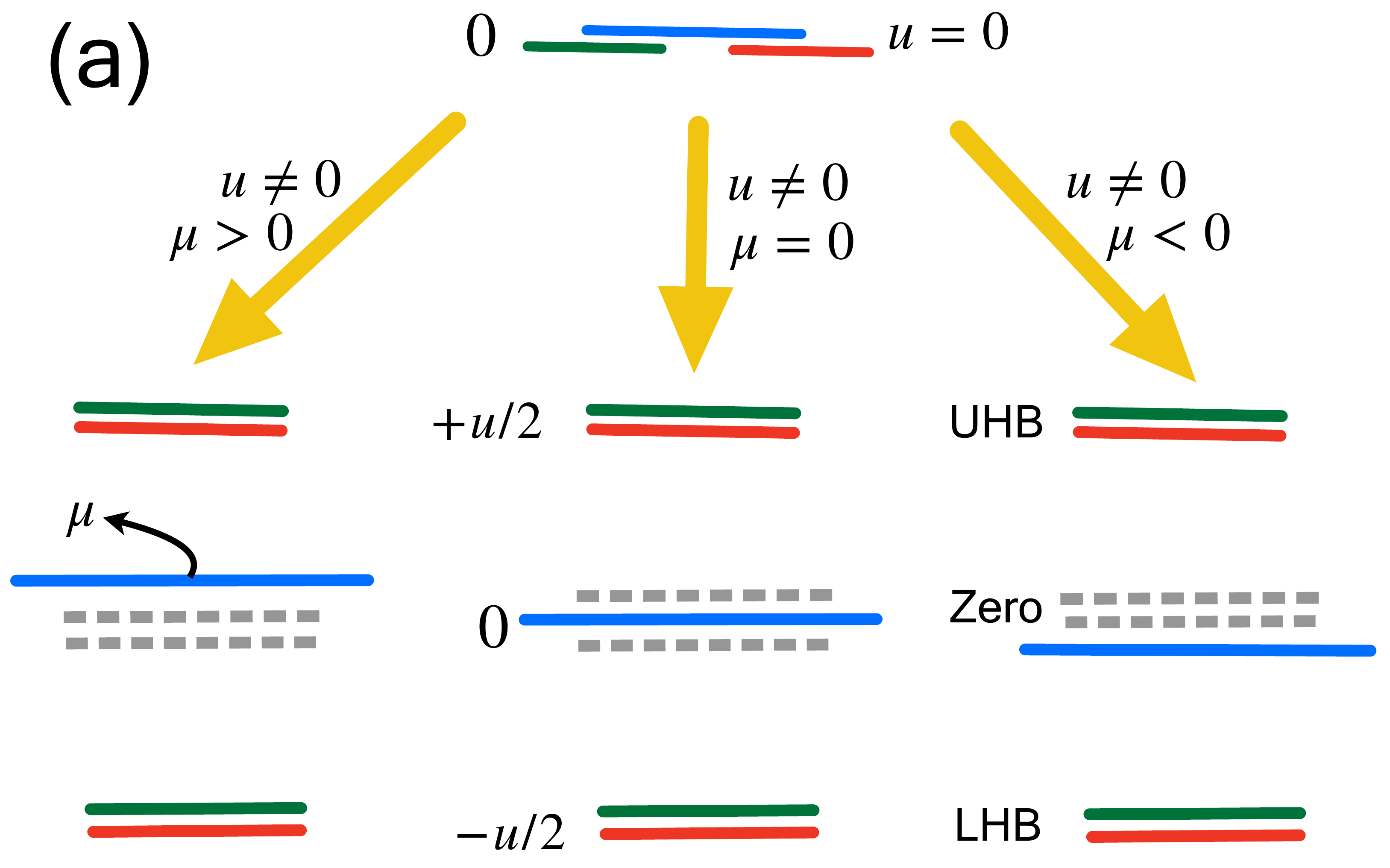} 
       \includegraphics[width=8cm, height=10cm,keepaspectratio]{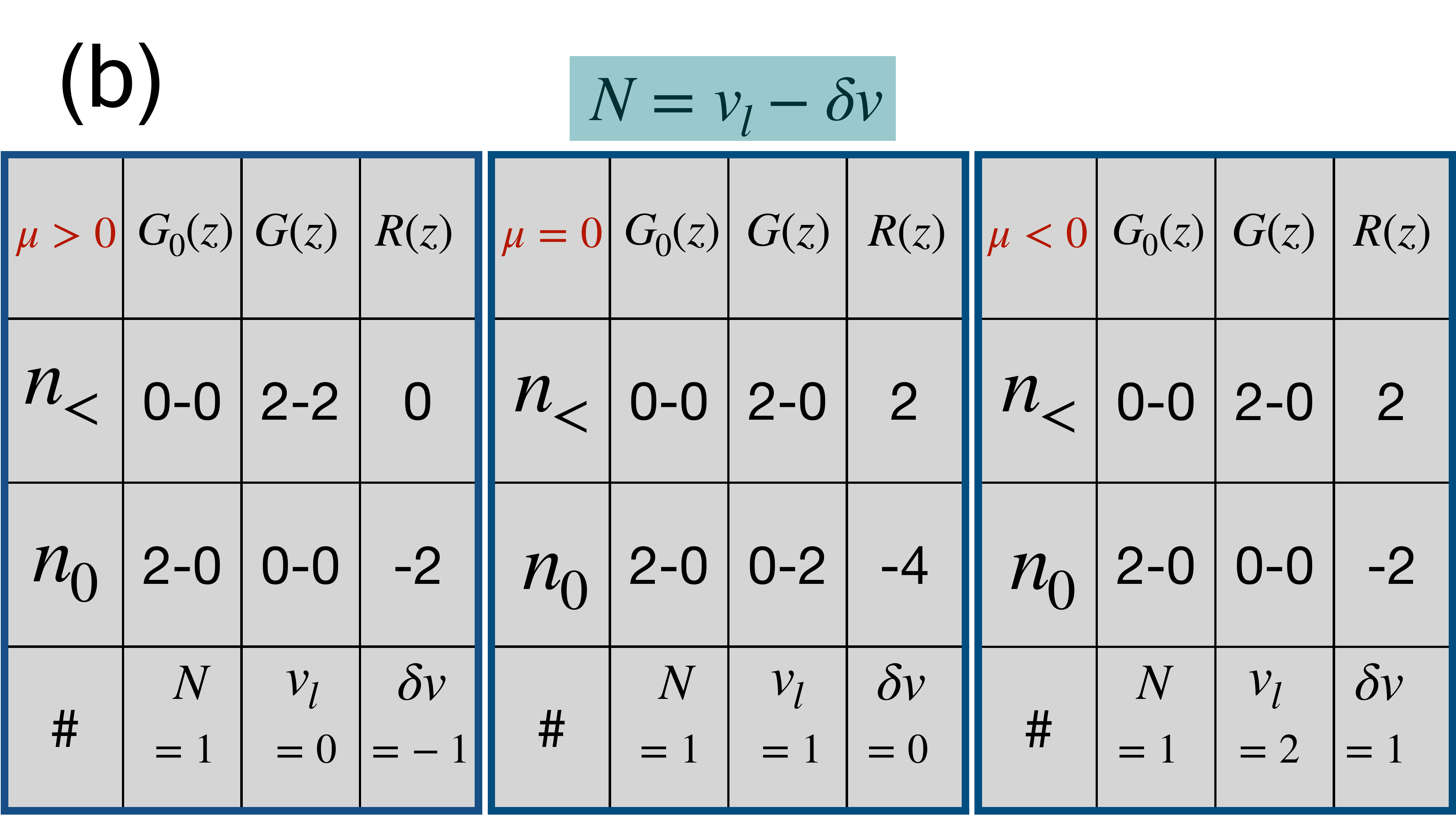} 
         \caption{ Illustration of Green's function zeros' contribution to the total particle count and failure of pole counting. (a)  
        Splitting of two occupied electronic states (red and green denote spin) at the Fermi energy into upper and lower Hubbard bands (UHB and LHB respectively) due to the
        repulsive Coulomb interaction $u$ for the cases of $\mu>0$, $\mu=0$ and $\mu<0$ in the atomic limit of the Hubbard model. The blue (dashed) line denotes the chemical potential (Green's function zeros). The vertical axis is frequency. 
         (b) Tables showing the difference between the numbers of poles and zeros (\# of poles - \# of zeros) for the three cases of $\mu>0$, $\mu=0$ and $\mu<0$. The rows $n_{<}(f)$,$n_0(f)$ label (\# of poles - \# of zeros) below and at the chemical potential respectively of the matrix $f$. ``$\#$" in the third row denotes the total contribution to the formula $N = v_l - \delta v$ in Eq.~\ref{ParticleNumberCount} using Eqs.~\ref{Eq:NSimplified}~\ref{Eq:vlSimplified}~\ref{Eq:dvSimplified}. The first two columns label the cases of non-interacting and interacting Green's function respectively and the third denotes the ratio of their determinants $R(z)$ (see Eq.~\ref{Eq:Rz}). In each of the three cases, $N = v_l - \delta v$ is satisfied and the particle number is conserved despite variations of the Luttinger volume $v_l$ and its backflow deviation $\delta v$. 
         }
     \label{Fig:Hubbard}
 \end{figure}

 
\textcolor{black}{More specifically, we compute here the charge and current response functions in accordance with the Ward identities and demonstrate how zeros manifest in physical observables. We begin by describing the exactly solvable models of interest in Sec.II (see Eqs.~\ref{Eq:MultibandHK}-~\ref{Eq:Hubbard}). In Sec.III, we reexamine the relationship between the Luttinger volume and self-energy in generic interacting settings. 
Using this relation, we provide a simple picture that elucidates how zeros are needed to preserve charge conservation while maintaining robustness of the total charge to changes in the chemical potential of the order of the Mott gap (see Figs.~\ref{Fig:Hubbard}, ~\ref{Fig:HK-Count}). 
More specifically, the total charge contains a term associated with the Luttinger volume and a ``backflow" term (see Eq.~\ref{ParticleNumberCount}).
We use our procedure of considering the total charge as a guide to analyze other more involved physical quantities, with a focus on the topological Hall response. In this way, 
in Sec. IV we evaluate the Hall response for a MI starting from non-interacting chern bands and show that it likewise
contains two contributions (see Eqs.~\ref{eqn:sigma-G-Lambda0},~\ref{eqn:N3expression},~\ref{eqn:sigma-xy-N3-relation}). The first is a quantized topological term proportional to the three dimensional winding number $N_3$~\cite{TKNN1982, Niu1985} with contributions from Green's function zeros (see Fig.~\ref{Fig:N3Variation}). The second is a previously unrecognized non-quantized backflow term essential to preserve charge conservation. 
In each case,
the two terms
combine to ensure that the 
total quantity is independent of changes to the chemical potential within the Mott gap \textit{despite} containing contributions from zeros. We discuss some implications and conclude in Sec. V.
}
\par
\section{Model}
 As 
 an
 exactly solvable model where Green's function zeros occur, we consider a generic multi-band version of the Hatsugai-Kohmoto (HK) model~\cite{HK1992}. We write the total Hamiltonian as
 \beq  \label{Eq:MultibandHK}
    H &=& H_0 + H_I \\
    H_0 &=&  \sum_{\bs k, \alpha\sigma,\beta\sigma'}h_{\alpha\sigma,\beta\sigma'}(\bs k)c^\dagger_{\bs k \alpha\sigma}c_{\bs k\beta\sigma'} \\
    H_I &=& \frac{U}{2}\sum_{\bs k \alpha} (n_{\bs k\alpha\uparrow} + n_{\bs k\alpha\downarrow} - 1)^2\,.
\eeq
Here $c_{\bs k\alpha \sigma}^{\dagger}$ is the electron creation  operator at momentum $\bs k$, orbital $\alpha$ and spin $\sigma$. The hopping matrix elements between states with orbital indices $\alpha, \beta$ and spin indices $\sigma, \sigma'$ are denoted by $h_{\alpha\sigma,\beta\sigma'}(\bs k)$, $U$ is a four-fermion electron-electron interaction that is local in momentum space but highly non-local in real space, and $n_{\bs k \alpha \sigma}$ is the number operator. 
The Hamiltonian at each $\vk$ point becomes mutually decoupled in the form $H = \sum_\vk H_\vk$, because of the local-in-momentum-space interaction.
Later in the paper, we will also use  a single band version of Eq.~\ref{Eq:MultibandHK} by replacing the kinetic hopping matrix by a single dispersion $\xi(\bs k) = \epsilon(\bs k) - \mu$ where $\epsilon(\bs k)$ is the band energy and $\mu$ the chemical potential. The interaction Hamiltonian then contains a single repulsive term between opposite spins at a specific momentum point. Accordingly, the orbital indices $\alpha, \beta$ are suppressed in the electron creation and annihilation operators in Eq.~\ref{Eq:MultibandHK} for the one band model.
We will explicitly specify when this is the case.  
The electronic and spectral properties of the Hamiltonian Eq.~\ref{Eq:MultibandHK} for single~\cite{HK1992, Setty2020, Setty2021, Setty2021-Kondo, Setty2018, Setty2023} and multi-band dispersions~\cite{Setty2023, Phillips2023, Bradlyn2023} have been previously studied but we recall some key properties. First, irrespective of the specific tight binding Hamiltonian at hand, Eq.~\ref{Eq:MultibandHK} captures a correlated metal to a fully gapped Mott insulator transition for interaction strength $U$ comparable or larger than the non-interacting bandwidth ($W$). Second, the Green's function can be obtained exactly, and in the limit of strong interactions compared to $W$ and zero temperature, a key property of the Green's function is the existence of contours of dispersive zeros in the Mott gap. These contours are a consequence of destructive cancellation of electron addition- and removal-like transitions with equal and opposite energy transfers~\cite{Setty2023}. Further,  lattice symmetries of the Hamiltonian $H$ constrain spectral degeneracies at high symmetry points that operate on both poles and zeros of the Green's function. Hence, Eq.~\ref{Eq:MultibandHK} offers a platform to explore topological properties in the presence of interactions non-perturbatively even when there is a loss of quasiparticles. Additional properties of $H$ for the case when the non-interacting bands have a non-trivial spin-Hall Chern number are discussed in Ref.~\cite{Setty2023}. Owing to these features, we use Eq.~\ref{Eq:MultibandHK} as a starting point for our analysis. \par
In section~\ref{Sec:TotalCharge} and associated Fig.~\ref{Fig:HK-Count}, it will suffice for us to work with a single band version of Eq.~\ref{Eq:MultibandHK}. We will use a quadratic band dispersion of the form $\xi (\bs k) = k^2 - \mu$ where we denote $k$ as the magnitude of $\bs k$. In section~\ref{Sec:Conductivity}, we will work with a multi-orbital tight-binding model with chern bands to compute the conductivity. In this case, the matrix elements $h_{\alpha\sigma,\beta\sigma'}(\bs k)$ are defined later in Eq.~\ref{Eq:TBModel}. \par
In the course of our discussion, we will also use the atomic limit of the Hubbard model to illustrate common features of certain conclusions. To establish notation, let us denote $n_{i\sigma}$ as the number operator at real space site $i$ and spin $\sigma$, and 
$u$ as the onsite coulomb interaction. In this limit, we write the Hubbard Hamiltonian as
\beq \label{Eq:Hubbard}
H_u = u \sum_i n_{i\uparrow} n_{i \downarrow} - \mu\sum_{i\sigma} c_{i\sigma}^{\dagger} c_{i\sigma}
\eeq
where $\mu$ is the onsite chemical potential, $c_{i\sigma}^{\dagger} $ is the electron creation operator at site $i$ and spin $\sigma$. In this limit, one obtains an atomic Mott insulator where the single site Green's function acquires dispersionless zeros when the probe frequency equals negative of the chemical potential~\cite{Fazekas1999}.  Thus certain aspects and properties of the Hamiltonian Eq.~\ref{Eq:MultibandHK} can be bench marked against the physics of the Hubbard model. We now examine how physical properties are affected by the presence of zeros in the Green's function in Eqs.~\ref{Eq:MultibandHK} and \ref{Eq:Hubbard}. 
\section{Total charge} \label{Sec:TotalCharge}
Before we study the specific case of Eq.~\ref{Eq:MultibandHK}, we begin by expressing the total particle number in terms of Green's function singularities~\cite{Yunoki2017}.  With knowledge of the interacting Green's function $G(z)$, the total particle number $N$ can be determined from the following equation ~\cite{Stefanucci2013, Mahan2000}:
\beq
N = \frac{1}{\beta} \sum_{\omega_n} \Tr[G(i\omega_n)] e^{i \omega_n \eta} = \oint \frac{dz}{2\pi i}n_F(z)\Tr[G(z)] \,,
\eeq
where $\omega_n$ is the fermionic Matsubara frequency, $\beta$ the inverse temperature, $\eta$ is an infinitesimally small positive number, $n_F(z)$ is the Fermi function, and the contour of integration encloses the Matsubara frequencies along the imaginary axis. To extract the topological winding characteristics of the particle number~\cite{Yunoki2017}, we note that the total Green's function can be written in terms of the non-interacting Green's function $G_0(i\omega_n)$ and self-energy $\Sigma(i\omega_n)$  through the Dyson equation $G(i\omega_n)^{-1} = G_0(i\omega_n)^{-1} - \Sigma(i\omega_n)$. Using this, we can rewrite
\beq
G(z) = G(z) \frac{\partial G(z)^{-1}}{\partial z} + G(z) \frac{\partial \Sigma(z)}{\partial z},
\eeq
and as a result the particle number $N$ takes the form
\beq \nonumber
N &=& 
\oint \frac{d z}{2\pi i} n_F(z)\left[\frac{\partial \ln \text{det} G(z)^{-1}}{\partial z} + \Tr\left(G(z)\frac{\partial \Sigma(z)}{\partial z}\right) \right] \\ 
&\equiv & v_l - \delta v.
\label{ParticleNumberCount}
\eeq
The first term in Eq.~\ref{ParticleNumberCount} denoted $v_l$ is traditionally defined as the Luttinger volume, and the second backflow term in Eq.~\ref{ParticleNumberCount} denoted $\delta v$ is its deviation from the total particle number. The Luttinger theorem states that $\lim_{\beta \rightarrow \infty} N = \lim_{\beta \rightarrow \infty} v_l$ .  
The theorem holds when particle-hole symmetry is preserved and we will see below that away from the
particle-hole 
symmetric
filling, the Luttinger theorem can be violated and the backflow term $\lim_{\beta \rightarrow \infty} \delta v \neq 0$. To further simplify Eq.~\ref{ParticleNumberCount}, we utilize analytical properties of the Green's function. In particular, we note that the determinant of the single particle Green's function can be decomposed into products of poles and zeros \cite{Gurarie2011}
\beq
\text{det} G(z) = \frac{\prod_{i=1}^{n_Z} (z - \zeta_i)}{\prod_{i=1}^{n_P}(z - \pi_i)},
\eeq
where $\zeta_i$ ($n_Z$) and $\pi_i$ ($n_P$) are the locations (number) of zeros and poles of the Green's function determinant. Substituting the factorizaton into the Luttinger volume gives a finite temperature expression
\beq
v_l = \sum_{i=1}^{n_P} n_F(\pi_i) - \sum_{i=1}^{n_Z} n_F(\zeta_i).
\eeq
At zero temperature, each of the poles (zeros) below the Fermi energy contribute one (negative one) count to the Luttinger volume while those at the Fermi energy contribute a $\frac{1}{2} (-\frac{1}{2})$ count. Hence 
the Fermi function 
is reduced to the Heaviside step (Kronecker delta) function for energies below (at) the Fermi energy,
and we can
rewrite the Luttinger volume
as
\beq \nonumber
v_l &=& \left(\sum_i^{n_P} \Theta(-\pi_i) - \sum_i^{n_Z} \Theta(-\zeta_i)\right) \\
&+& \frac{1}{2} \left( \sum_i^{n_P} \delta_{0,\pi_i} - \sum_i^{n_Z} \delta_{0,\zeta_i} \right).
\label{Eq:LVolume}
\eeq
For notational simplicity and later discussions, we will denote the difference between the number of poles and zeros below [at] the chemical potential for the determinant of the matrix $f$ as $n_{<}(f)[n_0(f)]$. Thus the $v_l$ can be rewritten succinctly as 
 \beq
 v_l =  n_{<}(G) + \frac{1}{2} n_0(G). 
 \label{Eq:vlSimplified}
 \eeq
 
 Notice that in the scenario when there are no zeros in the Green's function, the Luttinger volume reduces to its well known expression but with half the contribution from poles at the Fermi energy when compared to those below as must be expected. We can similarly simplify the expression for the deviation from the Luttinger volume $\delta v$. Defining the ratio of the determinant of the non-interacting and interacting Green's function as 
 \beq
 R(z) = \frac{\text{det} G_0(z)}{\text{det} G(z)},
 \label{Eq:Rz}
 \eeq
 we can rewrite the deviation as an expression similar to the Luttinger volume but in terms of analytical properties of $R(z)$ \cite{Yunoki2017}. Choosing a contour of integration that encloses the Matsubara frequencies along the imaginary axis, we have the backflow term~\cite{Yunoki2017}
\beq \nonumber
\delta v &=& -\oint \frac{d z}{2\pi i} n_F(z)  \Tr\left(G(z)\frac{\partial \Sigma(z)}{\partial z}\right)  \\ 
&=& + \oint \frac{d z}{2\pi i} n_F(z) \frac{\partial \ln R(z)}{\partial z}.
\eeq
We now define $Z_i$ ($N_Z$) and $\Pi_i$ ($N_P$) as the locations (number) of zeros and poles of $R(z)$. We then obtain an expression for the backflow $\delta v$ similar to that of $v_l$ in Eq.~\ref{Eq:LVolume} as
\beq \nonumber
\delta v &=& \left(\sum_i^{N_P} \Theta(-\Pi_i) - \sum_i^{N_Z} \Theta(-Z_i)\right) \\
&+& \frac{1}{2} \left( \sum_i^{N_P} \delta_{0,\Pi_i} - \sum_i^{N_Z} \delta_{0,Z_i} \right), \\ \label{Eq:RVolume}
&=& n_{<}(R^{-1}) + \frac{1}{2} n_0(R^{-1}).
\label{Eq:dvSimplified}
\eeq

In non-interacting systems, $G_0(z) = G(z)$ and by definition $R(z) = 1$ leading to $\delta v =0$.  The total particle number is fixed by that of the non-interacting electron Green's function and  takes the form
\beq
N = n_{<}(G_0) + \frac{1}{2} n_0(G_0).
\label{Eq:NSimplified}
\eeq
In a Fermi liquid 
and related
phases, 
due to the absence of Green's function zeros in its determinant, there exists a one-to-one mapping between the poles of $G_0(z)$ and  $G(z)$. This is because the lack of zeros leaves the pole structure of the Green's function determinant intact. Hence the Luttinger theorem continues to be satisfied and $\delta v = 0$.  
We are now in a position to calculate the particle number for different cases of interest including the Hamiltonians described in Eqs.~\ref{Eq:MultibandHK} and \ref{Eq:Hubbard}. \par

\subsection{Insights from the atomic limit}

\textit{Failure of pole counting:} To elucidate the role of quasiparticle loss on the particle number, we begin with the atomic limit of the Hubbard model in Eq.~\ref{Eq:Hubbard}. We argue that in this limit, counting of poles is insufficient to capture the total particle count. This fact is already reflected in Eqs.~\ref{ParticleNumberCount},~\ref{Eq:LVolume},and ~\ref{Eq:RVolume}, but here we give a simplified picture to help demonstrate a key notion -- Green's function zeros in the Mott gap contribute to the total particle number while also keeping it  invariant under changes to the chemical potential up to order of the gap. We argue that this holds true for any physically measurable property. Later in the paper, we will further reiterate this simple principle in the context of a topological 
Hall
response in the Mott phase obtained from the Hamiltonian in Eq.~\ref{Eq:MultibandHK}. 
\par
We work with the Hubbard Hamiltonian by setting the kinetic energy and chemical potential to zero, i.e.,
\beq
H_u = u\sum_i n_{i\uparrow} n_{i \downarrow}
\label{Eq:AtomicHubbard}
\eeq
where $i$ runs over the various sites of the lattice. 
In the absence of interactions, the determinant of the non-interacting Green's function per site is given by $\det G_0(z) = \frac{1}{z^2}$ where the quadratic power in the denominator is due to spin degree of freedom. This leads to \textit{two} poles located exactly at zero energy as displayed in Fig.~\ref{Fig:Hubbard} (a) with a weight of one-half each. In the presence of interactions and particle-hole symmetry, the determinant of the interacting Green's function $G_u(z)$ per site is given by 
\beq
\det G_u(z) = \left( \frac{4 z}{4 z^2 -u^2} \right)^2 \, ,
\label{Eq:AtomicHubbardDetG}
\eeq
where again the overall quadratic power is from the spin degree of freedom. Due to the pole in the self-energy $\Sigma(z) = \frac{u^2}{4 z}$, the interacting Green's function has \textit{four} 
poles -- \textit{two} poles each above and below the chemical potential -- and \textit{two} zeros at the chemical potential. This is shown below the center arrow in Fig.~\ref{Fig:Hubbard}(a).  As a result, there is a doubling of the number of poles \textit{below} the Fermi energy each with a weight of unity. Hence counting poles by themselves in the atomic limit of the Hubbard model cannot be sufficient to account for a fixed total particle number. A comparison of the determinants of $G_0(z)$ and $G_u(z)$ readily lays out the reason for the failure of pole counting -- a singular self-energy (Green's function zero) changes the order of the Green's function pole structure. Thus, while accounting for singularities of Green's function in the Hubbard model, it is essential to count zeros for preservation of the total particle number~\cite{AGD}. \par 
 \begin{figure}[h!]
     \centering
     \includegraphics[width=1 \linewidth]{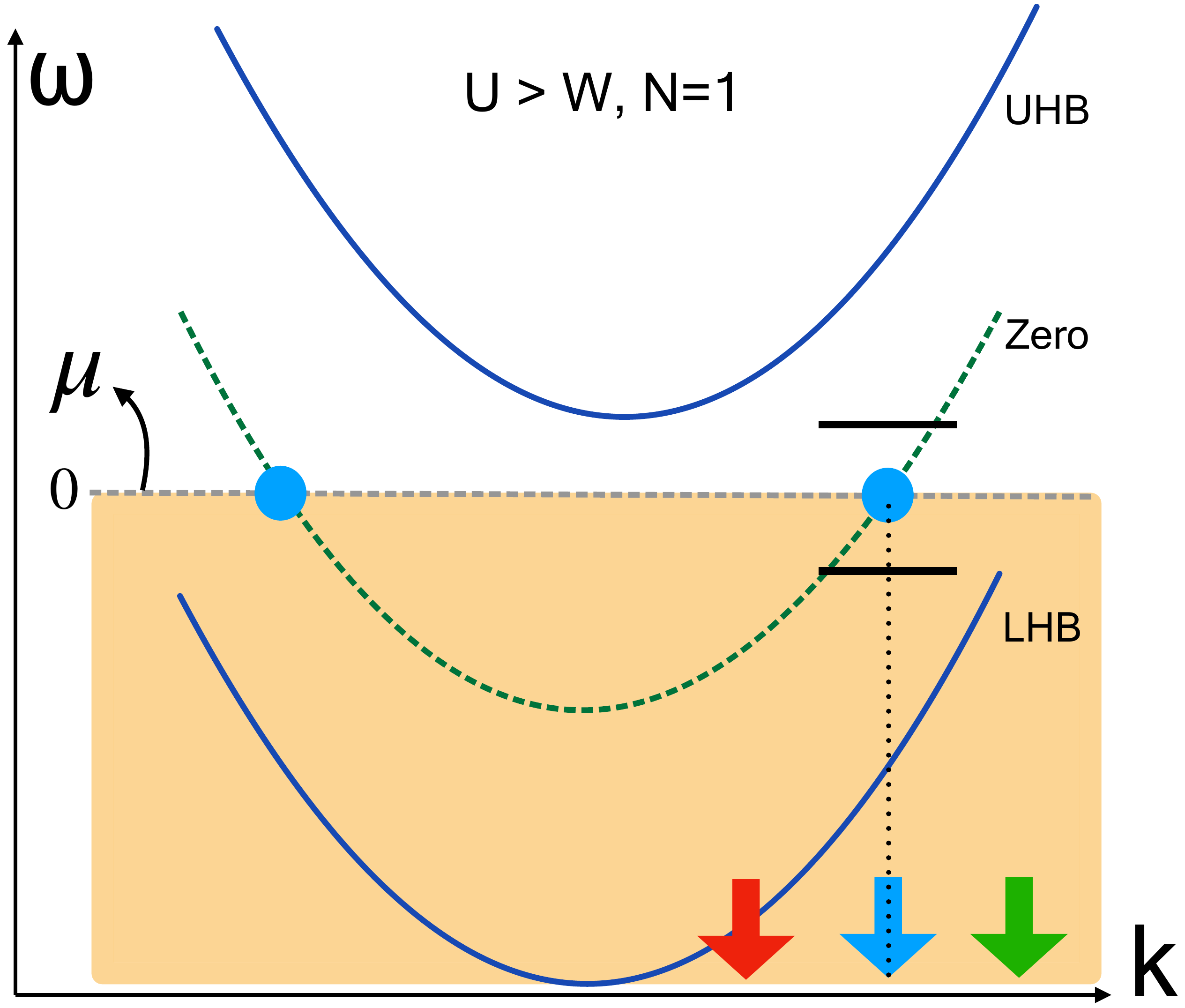}
         \caption{ Illustration of the upper Hubbard band (UHB), lower Hubbard band (LHB) and contour zero surface (dashed green line) of the model Hamiltonian in Eq.~\ref{Eq:MultibandHK} with a single band. The blue dots denote the intersection of zero surface with the Fermi energy (horizontal dashed gray line). The red (green) arrow denotes a $\bs k$ point within (outside) the zero surface. The blue arrow denotes a $\bs k$ point at the zero surface. The short solid lines above and below the right blue dot denote the cases when the chemical potential is moved slightly above and below the reference value. 
         }
     \label{Fig:HK-Count}
 \end{figure}
\textit{Role of zeros:} From dimensionality arguments, the singularity of the self-energy must naturally be involved to account for the total Luttinger count and electron number. Eqs. \ref{ParticleNumberCount}, ~ \ref{Eq:LVolume}, and~\ref{Eq:RVolume} precisely capture how Green's function zeros must be included to preserve the total particle number in the presence of interactions.
\par
To better clarify the role of zeros,  we compute the total particle number per site using Eqs. \ref{ParticleNumberCount}, ~ \ref{Eq:LVolume}, and~\ref{Eq:RVolume} in the atomic limit of the Hubbard model (Eq.~\ref{Eq:Hubbard}). We first work in the zero chemical potential limit and later consider scenarios where it is non-zero.  In the absence of interactions, we determine $v_l$ and $\delta v$ from the determinant of the non-interacting Green's function $\det G_0(z) = z^{-2}$. Since there are two poles at zero energy (Fig.~\ref{Fig:Hubbard} (a)), $v_l = \frac{1}{2}(2) = 1$, whereas $\ln R(z) = 0$ leading to $\delta v =0$; hence $N= v_l - \delta v = 1$ per site.  
In the presence of interactions, there are two poles below the Fermi energy and two zeros at the Fermi energy in the determinant Eq.~\ref{Eq:AtomicHubbardDetG} (Fig.~\ref{Fig:Hubbard} (a)). We therefore have $v_l = 2 - \frac{1}{2}(2) = 1$.  Whereas, since $R(z)^{-1} = \text{det} G_u(z) / \text{det} G_0(z) = \frac{4 z^4}{(4 z^2 - u^2)^2} $, we have two poles below and four zeros at the Fermi energy giving $\delta v = 2 - \frac{1}{2}(4) = 0$; hence again $N=v_l - \delta v =1$ per site. Thus we see that in the atomic limit of the Hubbard model, the Luttinger theorem holds when $\mu =0$ (particle-hole symmetry), and to recover the correct particle number, Green's function zeros must contribute to the count. 
\par
Moving away from the $\mu = 0$ limit, 
we shift chemical potential within the Mott gap away from the particle-hole symmetric point. For the case when $\mu<0$ in the presence of interactions, there are two poles (no singularities) below (at) the chemical potential (Fig.~\ref{Fig:Hubbard} (a)).
Hence we see that the Luttinger count is $v_l = 2+ \frac{1}{2}(0) = 2$ whereas the deviation is $\delta v = 2 + \frac{1}{2}(-2) = 1$, hence satisfying the same particle number condition $N = v_l - \delta v =1$. Similarly when $\mu>0$, there are two poles and zeros each (no singularities) below (at) the chemical potential (Fig.~\ref{Fig:Hubbard} (a)). We therefore have $v_l = 0 + \frac{1}{2}0 =0$ and $\delta v = 0+ \frac{1}{2}(-2) = -1$ so that the particle number $N = v_l - \delta v =1$ continues to remain unchanged. A summary of these numerical evaluations for the three cases of $\mu =0, \mu>0, \mu<0$ appears in Fig.~\ref{Fig:Hubbard} (b).   \par
\subsection{The case of 
Hatsugai-Kohmoto model}
We can apply a similar analysis to the dispersive bands of the Hatsugai-Kohmoto model of Eq.~\ref{Eq:MultibandHK}. It is sufficient to consider a single-band version of the Hamiltonian to illustrate our results. Since the model is local in momentum space where the individual $\bs k$ points are decoupled from each other, every such momentum point can be viewed as a single site Hubbard model with a different onsite energy. Hence the results from previous paragraphs for the single site Hubbard model in the atomic limit can be utilized in a straightforward manner.
\par
We focus on the limit where the interaction strength $U$ is larger than the bandwidth with an occupation $N=1$ per spin (half-filling) and momentum point. Fig~\ref{Fig:HK-Count} shows a schematic of the spectral function in this limit. The solid lines are the upper and lower `Hubbard-like' bands and the dashed line is the contour of Green's function zeros. The blue dots mark the intersection of the zero surface with the chemical potential and the red (green) arrow denotes a momentum point inside (outside) the zero surface at zero energy (Luttinger surface, LS). At the particle hole symmetric point (solid blue dot), there exist two poles (two zeros) below (at) the chemical potential. Hence, like in the particle-hole symmetric case described earlier we have $N = v_l = 1, \delta v =0$. Away from particle-hole symmetry, the chemical potential can be moved above or below the zero frequency (short black solid lines) for a given momentum corresponding to the blue arrow. Alternatively, the momenta can occur inside or outside the LS. For the case when the chemical potential is above zero, or equivalently, the momentum point lies inside the LS as marked by the red arrow, there are two poles and zeros below the chemical potential. As a result, $v_l =0$ but $\delta v = -1$ so that the net particle count $N=1$ continues to be preserved. Similarly, when the chemical potential is below zero, or equivalently, the momentum point lies outside the LS as marked by the green arrow, there are two poles below the chemical potential and we obtain $v_l =2$ but $\delta v = 1$ so that the net particle count is again $N=1$. Therefore, the particle number remains conserved for each momentum as expected for the Hamiltonian in Eq.~\ref{Eq:MultibandHK} regardless of changes in chemical potential.\par
The result above reconciles two seemingly conflicting notions: (1) that chemical potential changes less than or order of the Mott gap are not expected to affect physical properties since the ground state is unchanged and (2) that only "occupied" singularities (zeros or poles) contribute to the total particle number. \textcolor{black}{From our analysis above, we can conclude that indeed Green's function zeros contribute to the total particle number while also keeping it invariant to chemical potential changes smaller than $U$. The reconciliation between (1) and (2) is made possible because properly accounting for zero contributions is necessary to obtain the correct invariant value, and 
any deviation of the topological quantity (here the Luttinger count $v_l$)  from this value due to chemical potential changes is offset by an opposite variation of a non-topological one (here the backflow deviation $\delta v$) in Eq.~\ref{ParticleNumberCount}. We will see below that a similar mechanism holds for the transverse Hall response function where the role of the Luttinger volume is played by the topological number $N_3$ [\textit{c.f.}  Eqs.~\ref{eqn:sigma-G-Lambda0},~\ref{eqn:N3expression},~\ref{eqn:sigma-xy-N3-relation}].}  \par
\section{Hall Conductivity} \label{Sec:Conductivity}
In this section, we 
consider the electron transport properties 
in the presence of the Green's function zeros using Kubo formula \cite{coleman2015introduction}.
\textcolor{black}{We reiterate that
our analysis here of the transverse Hall response is guided by the 
procedure we have used while considering the total charge in the previous section.}
\subsection{Current operators}
The (optical) conductivity can be evaluated by the correlation functions of current operators, which are derived from the continuity equations $\partial_t \rho + \nabla \cdot \bm{j} = 0$. Combining the Heisenberg equation of motion with the continuity equation, one is able to obtain the expression for the current operator as follows:
\begin{equation}
    \vq \cdot \bm{j}_{\vq} = [H, \rho_\vq]\,,
\end{equation}
in which the Fourier transformed density operator takes the following form:
\begin{equation}
    \rho_\vq = \frac{1}{\sqrt{N_L}}\sum_{\vk,\alpha\sigma}c^\dagger_{\vk+\frac{\vq}{2}\alpha\sigma}c_{\vk - \frac{\vq}{2}\alpha\sigma}\,.
\end{equation}
Here $N_L$ stands for the size of the lattice. For a non-interacting Hamiltonian, the current operator is indeed the velocity operator of fermions:
\begin{equation}
    \bm{J}_{\vq} = \frac{1}{\sqrt{N_L}} \sum_{\vk, \alpha\sigma\beta\sigma'} \bm{v}_{\alpha\sigma,\beta\sigma'}(\vk)c^\dagger_{\vk + \frac{\vq}{2}\alpha\sigma}c_{\vk - \frac{\vq}{2}\beta\sigma}\,,
\end{equation}
in which $\bm{v}_{\alpha\sigma,\beta\sigma'}(\vk) = \nabla_{\vk}h_{\alpha\sigma,\beta\sigma'}(\vk)$ is the velocity matrix. However, when the interaction Hamiltonian contains terms which cannot be written as the products of local density operators (such as HK Hamiltonian), it will also contribute to the total current operator, which we will denote as $\bm{J}'_\vq$:
\begin{equation}
    \vq \cdot \bm{J}'_\vq = [H_I, \rho_\vq]\,,
\end{equation}
and the total current is the summation of the two terms $\bm{j}_\vq = \bm{J}_\vq + \bm{J}'_{\vq}$. Parameters in both the kinetic and interacting Hamiltonians will be affected by the external electromagnetic field $\bm{A}_{-\vq}$ via Peierls substitution due to the gauge invariance, because the creation and annihilation operators in the interacting Hamiltonian are do not always locate on the same lattice sites in real space. As a consequence, the current operator $\bm{J}_\vq'$ originated from the interacting Hamiltonian is also coupled to the gauge field. Similarly, the diamagnetic current can also originate from the interaction. 

\subsection{Conductivity tensor}
With all these factors considered, the optical conductivity with imaginary frequency can be written as summation of the current-current susceptibility and the diamagnetic response from both $\bm{J}_\vq$ and $\bm{J}_\vq'$. More precisely, the conductivity tensor takes the following form:
\begin{equation}\label{eqn:kubo}
    \sigma_{ij}(\vq, i\Omega) = \frac{1}{\Omega}\left(\mathcal{D}_{ij} + \chi_{ij}(\vq, i\Omega) + \mathcal{D}'_{ij}(\vq) + X_{ij}(\vq, i\Omega)\right),
\end{equation}
in which $\mathcal{D}_{ij}$ and $\mathcal{D}'_{ij}(\vq)$ are the diamagnetic response tensors obtained from expanding the kinetic Hamiltonian and interaction Hamiltonian to the second order of $\bm{A}_\vq$, and the susceptibilities $\chi_{ij}$ and $X_{ij}$ are defined as follows:
\begin{align}
    \chi_{ij}(\vq, i\Omega) &= - \int d\tau\, e^{i\Omega \tau} \langle T_\tau j^i_{\vq}(\tau) J^j_{-\vq}(0) \rangle\,,\label{eqn:def-chi-ij}\\
    X_{ij}(\vq, i\Omega) &= -\int d\tau \, e^{i\Omega \tau} \langle T_\tau j^i_\vq(\tau) J^{\prime j}_{-\vq}(0) \rangle\,.
\end{align}
Since the the interaction Hamiltonian usually contains four-fermion terms, the susceptibility $X_{ij}(\vq, i\Omega)$ could contain correlation functions with more than 6 fermionic operators. Using Ward-Takahashi identity \cite{nozieres2018theory, Betbeder1966}, we are able to rewrite the susceptibility $\chi_{ij}$ together with the diamagnetic term $\mathcal{D}_{ij}$ as a 
charge-current susceptibility:
\begin{align}
    &\mathcal{D}_{ij} + \lim_{\vq \rightarrow 0}\chi_{ij}(\vq, \Omega) = -i \Omega \lim_{\vq \rightarrow 0}\frac{\partial}{\partial q_i} \chi_{0j}(\vq, \Omega)\,,\label{eqn:Dij-chi-ij-chi-0j}\\
    & \chi_{0j}(\vq, i\Omega) = -\int d\tau e^{i\Omega \tau}\langle T_{\tau}\rho_\vq(\tau) J^j_{-\vq}(0)\rangle\,.\label{eqn:def-chi-0j}
\end{align}
The derivation of this relationship can be found in App.~\ref{app:ward}. Thus, the conductivity tensor will contain the 
charge-current susceptibility as follows:
\begin{align}
    & \sigma_{ij}(\vq \rightarrow 0, i\Omega) \nonumber \\
    = & -i\lim_{\vq\rightarrow 0} \frac{\partial}{\partial q_i}\chi_{0j}(\vq, i\Omega) + \sigma_{ij}'(\vq, i\Omega)\,,\label{eqn:sigma-chi0j-sigma-prime}
\end{align}
where $\sigma_{ij}'(\vq, i\Omega)$ stands for the contributions from $\mathcal{D}'_{ij}(\vq)$ and $X_{ij}(\vq, i\Omega)$, which contain correlation functions with 6 or more fermionic operators. 
In order to find the connection between the conductivity tensor and the Green's functions, it is better to write the charge-current susceptibility $\chi_{0j}$ as an integral of exact Green's functions $G(\vk, i\omega)$ and the exact vertex function $\Lambda^0(\vq, i\Omega; \vk, i\omega)$
(the definition of which
can be found in App.~\ref{app:ward}):
\begin{align}
    &\chi_{0j}(\vq, i\Omega) = \frac{1}{\sqrt{N_L}}\sum_\vk \int\frac{d\omega}{2\pi} {\rm Tr}\left[ G\left(\vk+\frac{\vq}{2},i\omega\right) \right. \nonumber\\ 
    &\cdot \left.\Lambda^0(\vk, i\omega; \vq, i\Omega) G\left(\vk - \frac{\vq}{2}, i\omega + i\Omega\right) v^j(\vk)\right]\,.
\end{align}
Taking the derivative of the susceptibility $\chi_{0j}$ with respect to the wave vector $\vq$, we yield the following expression for the conductivity tensor:
\begin{widetext}
\begin{align}
    \sigma_{ij}(i\Omega) =& -i\frac{1}{\sqrt{N_L}}\sum_\vk\int\frac{d\omega}{2\pi}{\rm Tr}\left[ G(\vk, i\omega) \frac{\partial \Lambda^0(\vk, i\omega;\vq, i\Omega)}{\partial q_i}\Big{|}_{\vq\rightarrow0}G(\vk, i\omega + i\Omega)v^j(\vk)\right]\nonumber\\
    &- \frac{i}{2\sqrt{N_L}}\sum_\vk\int\frac{d\omega}{2\pi}{\rm Tr}\left[\frac{\partial G(\vk, i\omega)}{\partial k_i} \Lambda^0(\vk, i\omega; \vq \rightarrow 0, i\Omega)G(\vk, i\omega + i\Omega)v^j(\vk)\right]\nonumber\\
    & + \frac{i}{2\sqrt{N_L}}\sum_{\vk}\int \frac{d\omega}{2\pi}{\rm Tr}\left[G(\vk, i\omega)\Lambda^0(\vk, i\omega;\vq \rightarrow 0,i\Omega)\frac{\partial G(\vk, i\omega + i\Omega)}{\partial k_i}v^j(\vk)\right] + \sigma_{ij}'(\vq, i\Omega)\,.
    \label{Eq:sigmaij}
\end{align}
One can easily notice that the second and 
third terms only contain the vertex function $\Lambda^0$ at $\vq \rightarrow 0$. Using the Ward-Takahashi identity for $\Lambda^\mu$, we are able to solve the vertex function $\Lambda^0(\vk, i\omega; 0,i\Omega)$ as follows:
\begin{equation}
    -i\Omega\cdot \Lambda^0(\vk, i\omega; \vq \rightarrow0, i\Omega) = \frac{\left[G^{-1}(\vk, i\omega) - G^{-1}(\vk, i\omega + i\Omega)\right]}{\sqrt{N_L}} + \lim_{\vq \rightarrow 0} \sum_i q_i \Lambda^i(\vq, i\Omega; \vk, i\omega)\,.
\end{equation}
Here the second term will vanish if the vertex functions $\Lambda^i$ do not diverge at $\vq = 0$. In normal condensed matter systems with short range interaction,
this condition is usually satisfied.
Because the HK model has 
a long-range interaction, we have, for generality, kept this term in the consideration.
In 
the DC limit $i\Omega \rightarrow 0$, it can also be written as:
\begin{align}
    \Lambda^0(\vk, i\omega; \vq \rightarrow0, i\Omega\rightarrow 0) =& -i\frac{1}{\sqrt{N_L}}\frac{\partial G^{-1}(\vk, i\omega)}{\partial \omega}  + \mathcal{F}(\vk, i\omega)\,,\label{eqn:vertex0-zerofreq-zeroq} \\
    \mathcal{F}(\vk, i\omega) = & \lim_{\Omega\rightarrow 0} \lim_{\vq \rightarrow 0}\sum_i \frac{q_i}{-i\Omega} \Lambda^{i}(\vq, i\Omega; \vk, i\omega)\,.
\end{align}
Using Eq.~\ref{eqn:vertex0-zerofreq-zeroq}, the conductivity at DC limit can be written as:
\begin{align}
    \sigma_{ij} &= -i\frac{1}{\sqrt{N_L}}\sum_\vk\int\frac{d\omega}{2\pi}{\rm Tr}\left[G(\vk, i\omega)\frac{\partial \Lambda^0(\vk, i\omega; \vq, i\Omega\rightarrow0)}{\partial q_i}\Big{|}_{q_i\rightarrow 0}G(\vk, i\omega)v^j(\vk)\right]\nonumber\\
    & - \frac{i}{2\sqrt{N_L}}\sum_\vk\int\frac{d\omega}{2\pi}{\rm Tr}\left[ \frac{\partial G(\vk, i\omega)}{\partial k_i} \mathcal{F}(\vk, i\omega) G(\vk, i\omega)v^j(\vk) \right] \nonumber\\
    & + \frac{i}{2\sqrt{N_L}}\sum_\vk\int\frac{d\omega}{2\pi}{\rm Tr}\left[ \partial G(\vk, i\omega) \mathcal{F}(\vk, i\omega) \frac{\partial G(\vk, i\omega)}{\partial k_i}v^j(\vk) \right] \nonumber\\
    & - \frac{1}{2N_L}\sum_\vk\int\frac{d\omega}{2\pi}{\rm Tr}\left[ \frac{\partial G(\vk, i\omega)}{\partial k_i} \frac{\partial G^{-1}(\vk, i\omega)}{\partial \omega} G(\vk, i\omega)v^j(\vk) \right]\nonumber\\
    & + \frac{1}{2N_L}\sum_\vk\int\frac{d\omega}{2\pi} {\rm Tr}\left[G(\vk,i\omega)\frac{\partial G^{-1}(\vk, i\omega)}{\partial \omega} \frac{\partial G(\vk, i\omega)}{\partial k_i}v^j(\vk)\right] + \sigma_{ij}'(\vq, i\Omega)\,.\label{eqn:sigma-Lambda0}
\end{align}
In addition to the $\sigma'_{ij}$ term from higher points correlation functions, 
Eq.~\ref{eqn:sigma-Lambda0} contain the information of  the vertex functions $\Lambda^\mu$ at 
nonzero $\vq$, which are inherently multi-point correlation functions and \emph{cannot} be represented by Green's functions.
\end{widetext}

\subsection{\texorpdfstring{$N_3$}{N3} and Hall conductivity}
We focus on the Hall conductivity $\sigma_{xy}$ in this paragraph. By using the identities $\partial G = -G\partial G^{-1} G$ and $v^j(\vk) = -\partial_{k_j}G^{-1}(\vk, i\omega) - \partial_{k_j}\Sigma(\vk, i\omega)$, we are able to represent the velocity matrices and the derivatives of the Green function as the derivatives of Green's function inverse matrices. Thus, the expression of the Hall conductivity can be rewritten as \textcolor{black}{(in analogy to Eq.~\ref{ParticleNumberCount} for the total particle number) }
\sur{
\begin{align}
    \sigma_{xy} =& N_3 + \Delta N_3,
    \label{eqn:sigma-G-Lambda0}
\end{align}
where the topological invariant 
\begin{widetext}
\begin{align}
N_3 =& \frac{1}{2N_L}\sum_{\vk}\int\frac{d\omega}{2\pi}{\rm Tr}\left[{ G(\vk,i\omega)\partial_{\omega}G^{-1}(\vk,i\omega)G(\vk,i\omega)\partial_{k_x} G^{-1}(\vk,i\omega)G(\vk,i\omega)}{ \partial_{k_y}G^{-1}(\vk, i\omega)}\right]\nonumber\\
    & -\frac{1}{2N_L}\sum_\vk\int\frac{d\omega}{2\pi}{\rm Tr}\left[ { G(\vk,i\omega)\partial_{k_x}G^{-1}(\vk, i\omega) G(\vk, i\omega)\partial_\omega G^{-1}(\vk, i\omega)G(\vk, i\omega)}{ \partial_{k_y}G^{-1}(\vk, i\omega)} \right] 
    \label{eqn:N3expression}
\end{align}    
\end{widetext}
\sur{gives the Hall conductivity in the non-interacting limit.
The difference,}
\begin{widetext}
\begin{align}
 \Delta N_3 =&  - i\frac{1}{\sqrt{N_L}}\sum_\vk\int\frac{d\omega}{2\pi}{\rm Tr}\left[G(\vk, i\omega)\frac{\partial \Lambda^0(\vk, i\omega; \vq, i\Omega\rightarrow0)}{\partial q_x}\Big{|}_{q_x\rightarrow 0}G(\vk, i\omega)v^y(\vk)\right]\nonumber\\
    & - \frac{i}{2\sqrt{N_L}}\sum_\vk\int\frac{d\omega}{2\pi}{\rm Tr}\left[ \frac{\partial G(\vk, i\omega)}{\partial k_i} \mathcal{F}(\vk, i\omega) G(\vk, i\omega)v^j(\vk) \right] \nonumber\\
    & + \frac{i}{2\sqrt{N_L}}\sum_\vk\int\frac{d\omega}{2\pi}{\rm Tr}\left[ \partial G(\vk, i\omega) \mathcal{F}(\vk, i\omega) \frac{\partial G(\vk, i\omega)}{\partial k_i}v^j(\vk) \right] \nonumber\\
    &+ \frac{1}{2N_L}\sum_{\vk}\int\frac{d\omega}{2\pi}{\rm Tr}\left[{ G(\vk,i\omega)\partial_{\omega}G^{-1}(\vk,i\omega)G(\vk,i\omega)\partial_{k_x} G^{-1}(\vk,i\omega)G(\vk,i\omega)}\partial_{k_y}\Sigma(\vk, i\omega)\right]\nonumber\\
    & -\frac{1}{2N_L}\sum_\vk\int\frac{d\omega}{2\pi}{\rm Tr}\left[ { G(\vk,i\omega)\partial_{k_x}G^{-1}(\vk, i\omega) G(\vk, i\omega)\partial_\omega G^{-1}(\vk, i\omega)G(\vk, i\omega)}\partial_{ky}\Sigma(\vk, i\omega) \right]\nonumber\\
    & + \sigma_{ij}'(\vq, i\Omega), 
\label{eqn:sigma-xy-N3-relation}
\end{align}
\end{widetext}
is an analog to $\delta v$ in the generalized Luttinger's theorem, and it vanishes in the absence of interactions.
Specifically,} in the non-interacting limit, the self energy $\Sigma(\vk, i\omega)$ is zero,  and the vertex function is a constant matrix $\Lambda^0(\vk, i\omega; \vq, i\Omega) = \mathds{1}/\sqrt{N_L}$ because of \emph{Wick's theorem} (instead of a consequence of Ward-Takahashi identity). 
The contribution from 6 or more point correlation functions $\sigma_{ij}'(\vq, i\Omega)$ also vanishes since $\bm{J}'_\vq = 0$. 
\sur{Thus, all the terms in the expression of $\Delta N_3$  are zero}, which indicates $\sigma_{xy} = N_3$. 
However, in the presence of interactions, the vertex $\Lambda^0$ is no longer a constant function of $\vq$, and the self-energy $\Sigma(\vk,i\omega) \neq 0$, and in general we
cannot identify the topological index $N_3$ as the Hall conductivity. The vertex functions $\Lambda^0$ inherently encodes a four-point function, and the Ward-Takahashi identity is only able to relate it to {\it both} the Green's functions and other vertex functions together, rather than representing everything solely through Green's function.

Nevertheless, both values of \( N_3 \) in \( \Delta N_3 \) receive contributions from the zeros of the Green's function, provided that interactions are properly accounted for. 
\sur{As exemplified by the Hatsugai-Kohomoto model, to be discussed below in Sec. IV D (see Fig.~\ref{Fig:N3Variation} and associated text), $N_3$ contains terms that are affected by the number of zeros below the chemical potential. Thus the choice of chemical potential even within an insulating gap can modify $N_3$ in close analogy to the Luttinger volume $v_l$ in Eq.~\ref{ParticleNumberCount}.  }
While the Green's function zeros' contributions to $N_3$ has been highlighted in Refs.~\cite{Gurarie2011,Phillips2023,Fabrizio2023}, we here establish such contributions in a consistent, gauge invariant way.
Therefore, caution must be exercised when considering Green's function zeros, particularly when computing observable quantities like \( \sigma_{xy} \) using Green's functions and other correlation functions. Despite this, later in the next section, we will show how the zero temperature conductivity remains unchanged under variations of the chemical potential within the Mott gap in spirit of our previous discussion on the total charge. 

A natural question regarding the Hall conductivity is whether $\sigma_{xy}$ will be equal to $N_3$ when $\Sigma(\vk, i\omega) = \Sigma(i\omega)$ is $\vk$ independent. Indeed, such self energy could eliminate the two terms in the fourth and fifth lines of Eq.~\ref{eqn:sigma-xy-N3-relation}, indicating that $\sigma_{xy}$ is differed from $N_3$ by only a term containing $\Lambda^0$ and $\sigma_{ij}'(\vq, i\Omega)$ which contains 6-point or 8-point correlation functions [when the vector vertex functions are well-behaved in the long-wavelength limit,
such that $\mathcal{F}(\vk, i\omega) = 0$]. One may wonder if the derivative of the vertex function $\Lambda^0$ can be represented by Green's functions via Ward-Takahashi identity. Since this remaining term contains the derivative of $\Lambda^0$ with respect to $\vq$, we are not able to use Eq.~\ref{eqn:vertex0-zerofreq-zeroq} directly. In fact, one could check the following ansatz vertex functions $\Lambda^\mu$ all satisfy the Ward-Takahashi identity:
\begin{align}
    &\Lambda^0(\vk,i\omega; \vq, i\Omega\rightarrow 0) \nonumber \\
    =& \frac{1}{\sqrt{N_L}}\left(\mathds{1} + i\frac{ \Sigma(i\omega + i\Omega) - \Sigma(i\omega)}{\Omega} - \vq \cdot \bm{\ell}\right)\,,\\
    &\Lambda^i(\vk, i\omega, \vq, i\Omega \rightarrow 0) = \frac{1}{\sqrt{N_L}} \left(v^i(\vk) + i\Omega \cdot \ell_i\right)\,.
\end{align}
Here $\bm{\ell}$ is a constant vector with the same dimension as length. The Ward-Takahashi identity is satisfied regardless of the choice of $\bm{\ell}$. Therefore, the derivative of the vertex function $\Lambda^0$ in the $\vq\rightarrow 0, \Omega \rightarrow 0$ limit will be:
\begin{equation}
    \frac{\partial \Lambda^0(\vk, i\omega; \vq, i\Omega\rightarrow 0)}{\partial q_x}\Big{|}_{q_x\rightarrow 0} = -\frac{\ell_x}{\sqrt{N_L}}\,.
\end{equation}
The value of $\ell_x$ is not able to be solely solved from the Ward-Takahashi identity. Thus, we cannot further reduce Eq.~\ref{eqn:sigma-xy-N3-relation} into an expression which only contains full Green's functions, even if the self energy $\Sigma$ is momentum $\vk$ independent. 

\subsection{HK model with Chern bands}\label{sec:hk-chern}

\begin{figure*}[t]
    \centering
    \includegraphics[width=.75\linewidth]{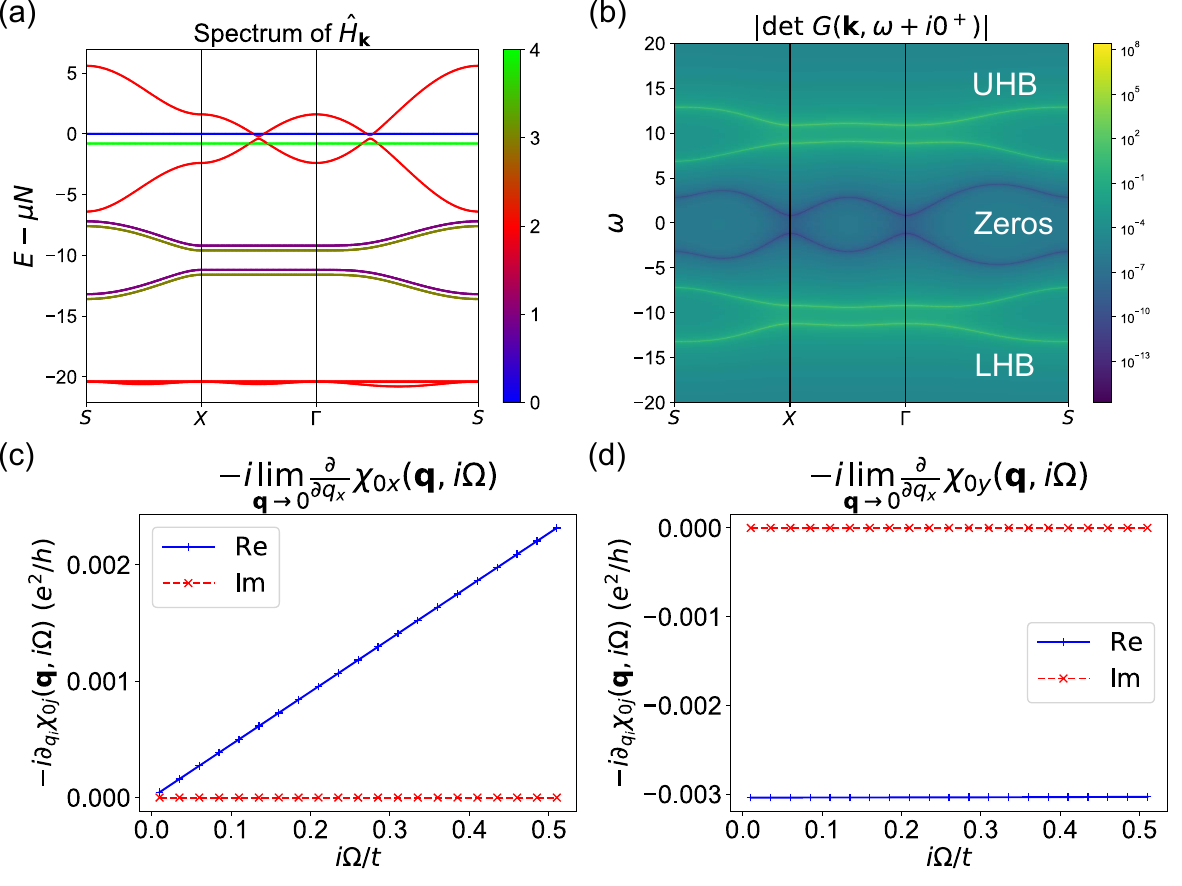}
    \caption{(a) The many-body energy spectrum of $H_\vk - \mu N_\vk$ along the high symmetry lines. Here we choose $t_{12} = t = 1$, $M = 1$ and $U = 20$. States with electron numbers $N_{\vk} = 0, 1, 2, 3, 4$ are labeled by blue, purple, red, brown and green respectively. In this figure we also added a chemical potential $\mu = 0.2$ to separate the $N_\vk = 1$ and $3$ states. (b) The Green's function determinant $|\det G(\vk, \omega+i0^+)|$ along the high symmetry lines. The corresponding value of $N_3$ computed from this Green's function is $N_3 = -2$. (c-d) Longitudinal and transverse components of the tensor $-i\lim_{\vq\rightarrow 0}\partial_{q_i}\chi_{0j}(\vq, i\Omega)$ as functions of imaginary frequency. The conductivity tensor $\sigma_{ij}$ is differed from this quantity by $\sigma_{ij}'(\vq, i\Omega)$ as shown in Eq.~\ref{eqn:sigma-chi0j-sigma-prime}. The real part value of $-i\lim_{\vq\rightarrow 0}\partial_{q_x}\chi_{0y}(\vq, i\Omega)$ at zero frequency is clearly different from the value of $N_3$, indicating that the backflow term $\Delta N_3$ 
    is nonzero.}
    \label{fig:chern-spectrum}
\end{figure*}

HK models are easily solvable using numerically exact diagonalization even if the kinetic energy $h_{\alpha\sigma,\beta\sigma'}(\vk)$ is not diagonal, due to the presence of a huge amount of good quantum numbers $N_\vk = \sum_{\alpha\sigma}c^\dagger_{\vk\alpha\sigma}c_{\vk\alpha\sigma}$. We choose a tight binding lattice model which carries a non-zero Chern number. The corresponding Hamiltonian is given by:
\begin{align}
    H_0 =& \sum_{\vk, \alpha\sigma,\beta\sigma'}h_{\alpha\sigma,\beta\sigma'}(\vk) c^\dagger_{\vk\alpha\sigma} c_{\vk\beta\sigma'}\,,\\
    h(\vk) =& \left[t_{12}\left(\tau_{1}\sin k_x + \tau_2\sin k_y \right) \right.\nonumber\\ \label{Eq:TBModel}
    & + \left.\tau_3(M - t \cos k_x - t\cos k_y)\right]\otimes s_0\,,\\
    H_I =& \frac{U}{2}\sum_{\vk}\sum_{\alpha=1}^2\left(n_{\vk\alpha\uparrow} + n_{\vk\alpha\downarrow} - 1\right)^2\,.
\end{align}
Here we use $\tau_{0, 1, 2, 3}$ to represent the identity and Pauli matrices with sublattice indices ($\alpha = 1,2$), and we use $s_{0,1,2,3}$ to represent the identity and Pauli matrices with spin indices ($\sigma = \uparrow, \downarrow$). Since the kinetic Hamiltonian $h(\vk)$ is proportional to $s_0$, it has a spin $SU(2)$ symmetry. When the parameters are chosen to be $t_{12} = t = 1$ and $|M| < 2$, the two energy bands of each spin will carry Chern numbers $\nu_C = \pm 1$, and due to the spin $SU(2)$ symmetry of the whole kinetic Hamiltonian, the two lower energy bands have the same Chern number. In the numerical calculation, we will choose $M = 1$.
\begin{figure}[h!]
     \centering
     \includegraphics[width=1 \linewidth]{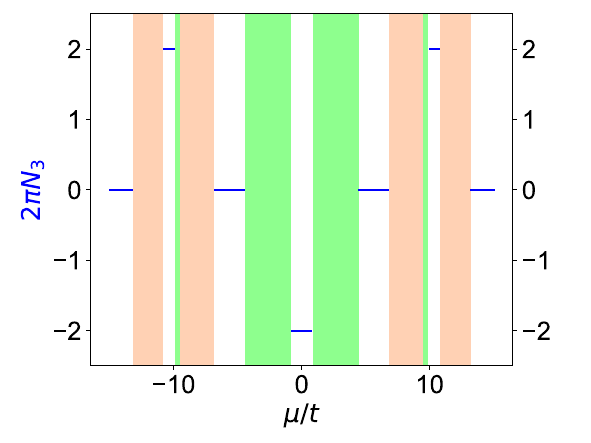}
         \caption{ \textcolor{black}{Variation of the topological winding number $N_3$ as a function of the chemical potential normalized by the hopping $t$. The zero (pole) bands are shown in green (orange) while the horizontal blue lines denote the values of $N_3$. $N_3$ is ill-defined in the energy windows where the zero and pole bands occur. The Mott gap is set to $20 t$.}
         }
     \label{Fig:N3Variation}
 \end{figure}
The Hamiltonian for every momentum value $H_\vk$ are completely decoupled from each other. Hence, the spectra and the wavefunctions for each $H_\vk$ can be numerically solved easily, since it is a $16\times 16$ matrix. In Fig.~\ref{fig:chern-spectrum}(a), we provide the energy spectra of $H_\vk - \mu N_\vk$ when the chemical potential is tuned such that the ground state is at half filling. We also note that the ground state for each $\vk$ will not change if the chemical potential is changed by a value $|\Delta \mu| \ll U/2$, due to the gap between the ground states ($N_\vk = 2$) and the charge $\pm 1$ ($N_\vk = 1,3$) excitations.

The many-body wavefunctions of the whole system are simply the tensor products of the wavefunctions for each $\vk$. With these exact wavefunctions in hand, we are able to compute quantities, such as Green's functions and susceptibilities via spectral decomposition. The determinant of the Green's function along high symmetry lines can be found in Fig.~\ref{fig:chern-spectrum}(b). Dispersive poles and zeros with different dispersion relationships are clearly visible. We also find numerically the topological index $N_3 = -2$ for this Green's function.

The derivative of the charge-current susceptibility $\chi_{0j}(\vq, i\Omega)$, which has been shown to be an important part of the 
conductivity tensor $\sigma_{ij}$, can also be evaluated numerically using spectral decomposition, as we discuss in App.~\ref{app:spec}. Figs.~\ref{fig:chern-spectrum}(c-d) provide the values of the $xx$ and $xy$ components of $-i\lim_{\vq\rightarrow 0}\partial_{q_i}\chi_{0j}(\vq, i\Omega)$ with imaginary frequencies. Clearly, the transverse component of this quantity is a non-zero small value. However, it is far from the value of $N_3$, \textcolor{black}{a plot of which appears in Fig.~\ref{Fig:N3Variation} for the HK model}. Thus, we can conclude that the
backflow terms $\Delta N_3$ in Eq.~\ref{eqn:sigma-xy-N3-relation} are not negligible even if $\sigma'_{xy}$ does not contribute.

We now comment on the extra term $\sigma'_{ij}(\vq, i\Omega)$ 
in the expression of the total conductivity tensor,
which describes the current $\bm{j}_\vq$ response from the coupling term between the external gauge field and the interaction current operator $\bm{J}_{-\vq}'$. 
As we have already mentioned, this term is inherently a 6-and-more-point correlation function, and it cannot be represented by the exact Green's function. 
It amounts to
an extra contribution to the backflow term $\Delta N_3$.

Finally,
we now show why the conductivity 
at zero temperature remains unchanged under variations in the chemical potential, provided that such changes do not alter the ground state. For
the
HK Hamiltonian at half filling, the ground states for each $\vk$ has a large gap $\sim U/2$ from any charge excitations, and as a consequence, the many-body ground states will remain in the half filling sector in a range of chemical potential. As we 
specified in Eq.~\ref{eqn:kubo}, the conductivity tensor contains the current-current susceptibility, which can be expressed 
in terms of a spectral decomposition:
\begin{align}
    &\chi_{ij}(\vq, i\Omega) + X_{ij}(\vq, i\Omega) \nonumber \\
    =& \frac{1}{|\mathcal{G}|}\sum_{g \in \mathcal{G}, m} \left[\frac{\langle g| j^i_{\vq}|m\rangle \langle m| j^j_{-\vq}|g\rangle}{i\Omega + (E_g - \mu N_g) - (E_m - \mu N_m)} \right.\nonumber\\
    &\left. - \frac{\langle g | j_{-\vq}^j| m\rangle \langle m | j^i_{\vq} | g\rangle}{i\Omega - (E_g - \mu N_g) + (E_m - \mu N_m)}\right]\,
    .
    \label{eqn:chi-spec-decomp}
\end{align} 
Here,
$\mathcal{G}$ denotes the set of degenerate ground states, and $|\mathcal{G}|$ stands for the ground states degeneracy. 
In addition,
$E_g, E_m$ and $N_g, N_m$ stand for the eigenvalues of the many-body Hamiltonian and the fermion number operator of many-body eigenstates $|g\rangle$ and $|m\rangle$, respectively. The current operators $\bm{j}_\vq$ always contain the same amount of fermion creation and annihilation operators. As a consequence, any excited state $|m\rangle$ which satisfies $\langle g | j^i_{\vq} | m\rangle \neq 0$ must have the same fermion number eigenvalue ($N_g = N_m$). Hence, the chemical potential $\mu$ does not show up in the denominator. If the ground states $|g\rangle$ remain unchanged when varying $\mu$ (which is true for a charge gapped system), Eq.~\ref{eqn:chi-spec-decomp} will also remain unchanged, regardless of whether the Green's function zeros are at the Fermi level or not.
\textcolor{black}{ Meanwhile, the diamagnetic response tensors $\mathcal{D}_{ij}$ and $\mathcal{D}'_{ij}(\vq)$ are directly determined by the ground state expectation values of fermionic operator products, which are 
unchanged if the ground state remains the same while changing $\mu$ even if it passes through zeros. As a consequence, the conductivity tensor will not be changed as well. This is despite the fact that (i)
the value of $N_3$ can be changed by varying chemical potential across zero bands. The variation of $N_3$ for the Hatsugai-Kohomoto model is shown numerically in Fig.~\ref{Fig:N3Variation}, and 
occurs even if the ground state is unaffected; and  (ii) $\sigma_{xy}$ must receive contributing terms from zeros that must be properly accounted for. We infer this from our numerical evaluation of $N_3$ for the HK model in Fig.~\ref{Fig:N3Variation} as well as  Eqs.~\ref{eqn:sigma-G-Lambda0},~\ref{eqn:N3expression},~\ref{eqn:sigma-xy-N3-relation})in analogy to the total charge in Eq.~\ref{ParticleNumberCount}. 
With these fully gauge invariant calculations, we have thus demonstrated how zero contributions to both the total charge and Hall conductivity can be properly accounted for, while at the same time, the physical observable itself remains invariant under chemical potential variations within the Mott gap.
We stress that our emphasis here is not the distinctness between the Hall conductivity and $N_3$ per se~\cite{Phillips2023} but on the behavior of the Hall conductivity itself.
}

\section{Discussion and Summary}
Several remarks are in order. First, as stated in the Introduction, probing Green's function zeros experimentally is not a straightforward task. Our objective in this paper is to study how zeros \textit{contribute} to certain observables in a consistent, gauge invariant manner. Extracting this contribution unambiguously is a challenge and a subject of ongoing debate which we postpone for a future investigation. A naive application of ordinary spectroscopic tools such as photoemission, tunneling or x-ray/neutron scattering
does not automatically reveal their existence and a more
nuanced approach is necessary. 
Instead, one might rely on specific probes that could extract this information more indirectly. For example, a key mechanism for the occurrence of zeros is through ground-state degeneracies, 
where we expect the zeros to come from a resonance scattering of electrons from singular collective spin excitations.
In this regard, a Curie-like behavior of the static, long wavelength magnetic susceptibility in the zero field limit with a Mott gap is a strong indicator of Green's function zeros. In principle, processes connecting non-degenerate but mixed ground states to excited states can also yield zeros at zero temperature. Such scenarios must be treated on a case-by-case basis.  Second, in the current paper we have not derived explicit forms of the 
six-and-more-point correlation
functions dictated by gauge invariance and alluded to in Eq.~\ref{Eq:sigmaij}. These terms 
contribute to the deviation of $\sigma_{xy}$ from its topological invariant $N_3$ as its particle number counterpart in Eq.~\ref{ParticleNumberCount}.
Nonetheless, we are able to reach the key conclusion 
that
measurable properties
are contributed to by Green's function zeros,
while remaining unchanged with chemical potential up to the Mott scale.
   \par
 In summary, in this work we have examined the role of  quasiparticle loss on  physical properties by studying an exactly solvable model of a Mott insulator.
 The model contains contours in momentum-frequency space where the Green's function vanishes to yield zeros within the Mott gap. We demonstrate that these zeros contribute to physical properties,
 such as the total particle number and conductivity tensor,
 in a way that is consistent with the expectations from general physical grounds. As an example of the latter,
 the observables are shown to be
 insensitive to changes in chemical potential within the Mott gap.
 Our results offer a conceptual framework for further analysis of topological response functions in strongly correlated systems and quantum materials where a well-defined quasiparticle picture is absent.
As such, we expect our work to help further advance the understanding of the rich interplay among topology, symmetry and strong correlations.

\textit{ Note Added:} After completing this manuscript, we became aware of a 
recently updated preprint and 
a new preprint, 
in which
the many-body effects 
of Hall conductivity
are also addressed \cite{Fabrizio2023, Goldman2023}.

\begin{acknowledgements}
We thank Jennifer Cano, Elio K\"onig, Diana-Gabriela Oprea, Silke Paschen and Roser Valenti for useful discussions.
Work at Rice has primarily been supported by the Air Force Office of Scientific Research under 
Grant No.
FA9550-21-1-0356 (C.S. and S.S.), 
by the National Science Foundation
under Grant No. DMR-2220603 (F.X. and L.C.),
and by the Robert A. Welch Foundation Grant No. C-1411 and the Vannevar Bush Faculty Fellowship ONR-VB N00014-23-1-2870 (Q.S.).
The majority of the computational calculations have been performed on the Shared University Grid at Rice funded by NSF under Grant EIA-0216467, a partnership between Rice University, Sun Microsystems, and Sigma Solutions, Inc., the Big-Data Private-Cloud Research Cyberinfrastructure MRI-award funded by NSF under Grant No. CNS-1338099, and the Extreme Science and Engineering Discovery Environment (XSEDE) by NSF under Grant No. DMR170109. 
M.G.V. acknowledges support of the Deutsche Forschungsgemeinschaft (DFG, German Research Foundation) 
GA 3314/1-1 - FOR 5249 (QUAST) and partial support from European Research Council (ERC) grant agreement no. 101020833. 
This work has also been funded by the European Union NextGenerationEU/PRTR-C17.I1, 
as well as by the IKUR Strategy under the collaboration agreement between Ikerbasque Foundation 
and DIPC on behalf of the Department of Education of the Basque Government.
All authors acknowledge 
the hospitality of the Kavli Institute for Theoretical Physics, UCSB,
supported in part
by the National Science Foundation under Grant No. NSF PHY-1748958,
 during the program ``A Quantum Universe in
a Crystal: Symmetry and Topology across the Correlation Spectrum."  Q.S. and S.S. also 
acknowledge the hospitality of the Aspen Center for Physics, which is supported by the National Science Foundation under Grant No. PHY-2210452, during the workshop ``New Directions on Strange Metals in Correlated Systems." C.S. acknowledges partial support from Iowa State University
startup funds.
\end{acknowledgements}

$\dagger$ \href{csetty@iastate.edu}{csetty@iastate.edu}
   
$\oplus$ \href{fx7@rice.edu}{fx7@rice.edu}

\onecolumngrid
\appendix

\section{Ward-Takahashi identity and susceptibilities}\label{app:ward}
In this appendix, we provide a detailed discussion regarding the susceptibilities $\chi_{ij}$, $\chi_{0j}$ and the relationship between these susceptibilities and the Ward identity \cite{nozieres2018theory, Betbeder1966}. 

The Ward-Takahashi identity can be formulated into the following form using imaginary time representation:
\begin{align}
    &\partial_\tau \Big{\langle} T_\tau \rho_\vq(\tau) c^\dagger_{\vk - \frac{\vq}{2} \alpha\sigma}(\tau') c_{\vk + \frac{\vq}{2}\beta\sigma'}(\tau'') \Big{\rangle} \nonumber\\
    =& \vq \cdot\Big{\langle} T_{\tau} \bm{j}_{\vq}(\tau) c^\dagger_{\vk - \frac{\vq}{2} \alpha\sigma}(\tau') c_{\vk + \frac{\vq}{2}\beta\sigma'}(\tau'')  \Big{\rangle}+ \frac{\delta(\tau' - \tau)}{\sqrt{N_L}} G_{\beta\sigma',\alpha\sigma}\left(\vk + \frac{\vq}{2}, \tau'' - \tau \right) - \frac{\delta(\tau'' - \tau)}{\sqrt{N_L}} G_{\beta\sigma',\alpha\sigma}\left(\vk - \frac{\vq}{2}, \tau - \tau'\right)\,,\label{eqn:ward-time-domain}
\end{align}
in which $G_{\beta\sigma',\alpha\sigma}(\vk, \tau) = -\langle T_\tau c_{\vk,\beta\sigma'}(\tau) c_{\vk,\alpha\sigma}^\dagger(0)\rangle$ represents the imaginary time Green's function. This identity can also be written in the frequency domain, by performing the Fourier transformation on both sides. A compact way to rewrite the frequency domain Ward-Takahashi identity is:
\begin{equation}\label{eqn:ward-Q}
    -i\Omega\cdot Q^0_{\alpha\sigma,\beta\sigma'}(\vk, i\omega; \vq, i\Omega) = \vq \cdot \bm{Q}_{\alpha\sigma, \beta\sigma'}(\vk, i\omega; \vq, i\Omega) + \frac{1}{\sqrt{N_L}}G_{\beta\sigma',\alpha\sigma}\left(\vk + \frac{\vq}{2}, i\omega\right) - \frac{1}{\sqrt{N_L}}G_{\beta\sigma',\alpha\sigma}\left(\vk -\frac{\vq}{2}, i\omega + i\Omega\right)\,,
\end{equation}
in which the correlation functions $Q^\mu_{\alpha\sigma,\beta\sigma'}(\vk, i\omega; \vq, i\Omega)$ are defined as:
\begin{align}
    Q^0_{\alpha\sigma,\beta\sigma'}(\vk, i\omega; \vq, i\Omega) &= \int d\tau \int d\tau'' e^{i(\Omega \tau + \omega\tau'')} \langle T_{\tau} \rho_\vq(\tau) c^\dagger_{\vk -\frac{\vq}{2}\alpha\sigma}(0) c_{\vk + \frac{\vq}{2}\beta\sigma'}(\tau'') \rangle\,,\\
    Q^i_{\alpha\sigma,\beta\sigma'}(\vk,i\omega; \vq, i\Omega) &= \int d\tau \int d\tau'' e^{i(\Omega \tau + \omega \tau'')} \langle T_{\tau} j^i_\vq(\tau) c^\dagger_{\vk -\frac{\vq}{2}\alpha\sigma}(0) c_{\vk + \frac{\vq}{2}\beta\sigma'}(\tau'') \rangle\,.
\end{align}

Another useful form of the Ward-Takahashi identity connects the Green's functions and vertex functions $\Lambda^\mu(\vq, i\Omega; \vk, i\omega)$, which are defined as:
\begin{equation}
    \Lambda^\mu(\vk, i\omega; \vq, i\Omega) = -G^{-1}\left(\vk + \frac{\vq}{2}, i\omega\right) \left[Q^\mu(\vk, i\omega; \vq, i\Omega)\right]^{\rm T}G^{-1}\left(\vk - \frac{\vq}{2}, i\omega + i\Omega\right)\,.
\end{equation}
Thus, we can obtain the Ward-Takahashi identity using the vertex functions by rewriting Eq.~\ref{eqn:ward-Q}:
\begin{equation}
    -i\Omega \cdot \Lambda^0(\vk, i\omega; \vq, i\Omega) = \sum_{i}q_i \cdot \Lambda^i(\vk, i\omega;\vq, i\Omega) - \frac{1}{\sqrt{N_L}}\left[G^{-1}\left(\vk -\frac{\vq}{2}, i\omega + i\Omega\right) - G^{-1}\left(\vk + \frac{\vq}{2}, i\omega\right)\right]\,.
\end{equation}

It is obvious that the susceptibilities $\chi_{ij}(\vq, i\Omega)$ and $\chi_{0j}(\vq, i\Omega)$, which are defined in Eq.~\ref{eqn:def-chi-ij} and Eq.~\ref{eqn:def-chi-0j}, can be represented by the correlation functions $Q^\mu(\vq, i\Omega; \vk, i\omega)$ in the following form:
\begin{align}
    \chi_{0j}(\vq, i\Omega) &= -\int d\tau e^{i\Omega\tau}\langle T_{\tau}\rho_\vq(\tau)J^j_{-\vq}(0)\rangle = -\frac{1}{\sqrt{N_L}}\sum_\vk\int\frac{d\omega}{2\pi} {\rm Tr}\left[\left(Q^{0}(\vk, i\omega;\vq, i\Omega)\right)^{\rm T}v^j(\vk)\right] \,,\\
    \chi_{ij}(\vq, i\Omega) &= -\int d\tau e^{i\Omega\tau}\langle T_{\tau} j^{i}_{\vq}(\tau) J^j_{-\vq}(0) \rangle = -\frac{1}{\sqrt{N_L}}\sum_\vk\int\frac{d\omega}{2\pi} {\rm Tr}\left[\left(Q^{i}(\vk, i\omega;\vq, i\Omega)\right)^{\rm T}v^j(\vk)\right]\,.
\end{align}
Combining this with the Ward-Takahashi identity shown in Eq.~\ref{eqn:ward-Q}, we can find how these two susceptibilities are related to the Green's functions:
\begin{equation}\label{eqn:susceptibility-greensf-relation}
    -i\Omega\cdot \chi_{0j}(\vq, i\Omega) = \sum_i q_i \chi_{ij}(\vq, i\Omega) - \int_{\rm BZ}\frac{d^dk}{(2\pi)^d}\int\frac{d\omega}{2\pi}{\rm Tr}\left[\frac{\partial h(\vk)}{\partial k_j}\left(G\left(\vk + \frac{\vq}{2}, i\omega\right) - G\left(\vk - \frac{\vq}{2}, i\omega + i\Omega\right)\right)\right]\,.
\end{equation}
By taking the derivative with respect to $q_i$ and then taking the $\vq \rightarrow 0$ limit, we are able to obtain the expression for the $\chi_{ij}(\vq, i\Omega)$ as shown:
\begin{equation}\label{eqn:susceptibility-greensf-relation-derivative}
    \chi_{ij}(\vq\rightarrow 0, i\Omega) = -i\Omega \lim_{\vq \rightarrow 0} \frac{\partial}{\partial q_i}\chi_{0j}(\vq, i\Omega) + \int_{\rm BZ}\frac{d^dk}{(2\pi)^d}\int\frac{d\omega}{2\pi}{\rm Tr}\left[ \frac12 \frac{\partial h(\vk)}{\partial k_j} \frac{\partial}{\partial k_i}\Big{(} G(\vk, i\omega) + G(\vk, i\omega + i\Omega) \Big{)} \right]\,.
\end{equation}
If the integral over $\omega$ in the second term is {\it absolute convergent}, it will not depend on the value of $\Omega$, since it is equivalent to a shift of the variable of integration $\omega \rightarrow \omega + \Omega$. Hence, Eq.~\ref{eqn:susceptibility-greensf-relation-derivative} can be further simplified into:
\begin{equation}\label{eqn:ij-susceptibility-as-0j-GG}
    \chi_{ij}(\vq\rightarrow 0, i\Omega) = -i\Omega \lim_{\vq \rightarrow 0} \frac{\partial}{\partial q_i}\chi_{0j}(\vq, i\Omega) + \int_{\rm BZ}\frac{d^dk}{(2\pi)^d}\int\frac{d\omega}{2\pi}{\rm Tr}\left[ \frac{\partial h(\vk)}{\partial k_j} \frac{\partial G(\vk, i\omega)}{\partial k_i}\right]\,.
\end{equation}
Using this equation, we are able to show that Eq.~\ref{eqn:Dij-chi-ij-chi-0j} in the main text will hold if Ward-Takahashi identity is satisfied. The diamagnetic response tensor from the kinetic Hamiltonian $\mathcal{D}_{ij}$ has the following form:
\begin{equation}\label{eqn:diamagnetic-respose-tensor}
    \mathcal{D}_{ij} = \int_{\rm BZ}\frac{d^dk}{(2\pi)^d}\sum_{\alpha\sigma\beta\sigma'}\frac{\partial^2h_{\alpha\sigma,\beta\sigma'}(\vk)}{\partial k_i \partial k_j}\langle g | c^\dagger_{\vk\alpha\sigma}c_{\vk\beta\sigma'} | g \rangle\,.
\end{equation}
We can use the Green's functions to represent the fermion operator bilinear expectation values, which has been used in the proof of Luttinger theorem \cite{Yunoki2017}:
\begin{equation}
    \langle g | c^\dagger_{\vk\alpha\sigma}c_{\vk\beta\sigma'} | g \rangle = G_{\beta\sigma',\alpha\sigma}(\vk, \tau=0^-) = \int\frac{d\omega}{2\pi}G_{\beta\sigma',\alpha\sigma}(\vk, i\omega) e^{i\omega 0^+}\,,
\end{equation}
in which the factor $e^{i\omega 0^+}$ ensures that the integral for the diagonal elements $(\alpha\sigma = \beta\sigma')$ converge. By doing so, the diamagnetic response tensor becomes:
\begin{equation}
    \mathcal{D}_{ij} = \int_{\rm BZ}\frac{d^dk}{(2\pi)^d}\int\frac{d\omega}{2\pi}{\rm Tr}\left[\frac{\partial^2h(\vk)}{\partial k_i \partial k_j}G(\vk, i\omega)\right] e^{i\omega 0^+}\,.
\end{equation}
By performing the integration by parts with respect to $k_i$, this expression can further be transformed into:
\begin{equation}
    \mathcal{D}_{ij} = \int_{\rm BZ}\frac{d^dk}{(2\pi)^d}\int\frac{d\omega}{2\pi}{\rm Tr}\left[\frac{\partial}{\partial k_i}\left(\frac{\partial h(\vk)}{\partial k_j} G(\vk, i\omega)\right) - \frac{\partial h(\vk)}{\partial k_j}\frac{\partial G(\vk, i\omega)}{\partial k_i}\right]e^{i\omega0^+}\,.
\end{equation}
The first term, being a derivative of $k_i$, will vanish, since the kinetic Hamiltonian $h(\vk)$ and the interacting Green's function $G(\vk, i\omega)$ are periodic functions in the whole Brillouin zone. Therefore, only the second term remains. Because we have assumed that the integral over $\omega$ of $\partial_{k_i}G(\vk, i\omega)$ is absolute convergent, the infinitesimal exponential factor $e^{i\omega0^+}$ could be dropped. Thus, the diamagnetic response tensor exactly cancels the Green's function part in Eq.~\ref{eqn:ij-susceptibility-as-0j-GG}. We conclude that the summation of the diamagnetic response $\mathcal{D}_{ij}$ and the susceptibility $\chi_{ij}(\vq\rightarrow 0, i\Omega)$ can be written as:
\begin{equation}
    \mathcal{D}_{ij} + \chi_{ij}(\vq\rightarrow 0, i\Omega) = -i\Omega \lim_{\vq \rightarrow 0}\frac{\partial}{\partial q_i}\chi_{0j}(\vq, i\Omega)\,,\\
\end{equation}
which is indeed the form of Eq.~\ref{eqn:Dij-chi-ij-chi-0j} in the main text.

The above discussion is based on the assumption that the integral $\int d\omega \partial_{k_i}G(\vk, i\omega)$ is absolute convergent. Here we argue that this is indeed true using the analytic properties of Green's functions. The derivative of the Green's function with respect to momentum $k_i$ can be reexpressed as:
\begin{equation}
    \frac{\partial G(\vk, z)}{\partial k_i} = - G(\vk, z) \frac{\partial G^{-1}(\vk, z)}{\partial k_i} G(\vk, i\omega) = G(\vk, z)\left[v(\vk) + \frac{\partial \Sigma(\vk, z)}{\partial k_i}\right]G(\vk, z)\,,
\end{equation}
where $\Sigma(\vk, z) = z - h(\vk) - G^{-1}(\vk, z)$ is the self energy. In general, an element of the Green's function at large frequency decays as or faster than $\frac{1}{z}$, or more precisely, $\lim\limits_{z\rightarrow \infty} |z\cdot G_{\alpha\sigma,\beta\sigma'}(\vk, z)| \leq 1$. At high frequency, self energy becomes finite and frequency independent \cite{Luttinger1961}, as does its derivative $\lim\limits_{z \rightarrow \infty}|\partial_{k_i}\Sigma(\vk, z)| < \infty$. Therefore, each element of the derivative of the Green's function will satisfy the inequality:
\begin{equation}
    \lim_{z\rightarrow \infty}\Bigg{|}z^2 \frac{\partial G_{\alpha\sigma,\beta\sigma'}(\vk, z)}{\partial k_i} \Bigg{|} < \infty\,.
\end{equation}
Since the absolute value of the integrand decays as or even faster than $\frac{1}{\omega^2}$, the integral over $\omega$ in Eq.~\ref{eqn:susceptibility-greensf-relation-derivative} is absolute convergent.

\section{Spectral decomposition of Green's function and susceptibilities in HK models}\label{app:spec}

Since the electron numbers at each momentum point $\vk$ are all conserved quantities, the many-body eigenstates wavefunctions can be built by the tensor products of eigenstates of all $H_\vk$. In the HK model with Chern bands defined in Eq.~\ref{Eq:TBModel}, we have two energy bands per spin ($N_\alpha = 2$), the Hilbert space dimension of each $H_\vk$ is $\rm dim = 2^{2N_\alpha} = 16$. 
Among these states, $N_\vk = 0, 4$ has $\binom{4}{0} = \binom{4}{4}= 1$ eigenstate; $N_\vk = 1,3$ has $\binom{4}{1} = \binom{4}{3} = 4$ eigenstates; and $N_\vk = 2$ has $\binom{4}{2} = 6$ eigenstates. 

Due to the $SU(2)$ spin rotation symmetry, the $6\times 6$ Hamiltonian $H_\vk$ in the $N_\vk = 2$ sector can be reduced to a block diagonal form. 
Among these diagonal blocks, the largest is $3\times 3$, which contains three $SU(2)$ singlet states. This means the analytic results of many-body wavefunctions, Green's functions, and susceptibilities can only be expressed by cubic roots, which will be highly complicated and not be able to offer substantial insights. 
In contrast to the complicated analytic expressions involved in obtaining many-body wavefunctions, the Hamiltonian $H_\vk - \mu N_\vk$ is relatively straightforward to solve numerically since it is a $16\times 16$ matrix for each $\vk$. 
We denote these 16 states by $s = 1, 2, \cdots 16$, whose energies increase with their indices ($E_{s=1}(\vk) \leq E_{s=2}(\vk) \leq\cdots \leq E_{s=16}(\vk)$).
We can also use a string of state indices $\{s_\vk\}$ to represent a many-body wavefunction and energy:
\begin{align}
    |m\rangle &= \bigotimes_{\vk}|s^m_{\vk}\rangle\,,\\
    E_m &= \sum_\vk E_{s^m_\vk}(\vk)\,.
\end{align}
Using these notations, the zero temperature Green's function can be solved via the spectral decomposition as follows:
\begin{equation}\label{eqn:greens-function-spec-decom}
    G_{\alpha\sigma,\beta\sigma'}(\vk, z) = \frac{1}{D_{\vk}}\sum_{s^g_{\vk}=1}^{D_{\vk}}\sum_{s_{\vk}^m=1}^{16}\left(\frac{\langle s_\vk^g | c_{\vk\alpha\sigma} | s_\vk^m \rangle\langle s_\vk^m |c^\dagger_{\vk\beta\sigma'}| s_\vk^g \rangle}{z + E_{s_\vk^g}(\vk) - E_{s_\vk^m}(\vk)} - \frac{\langle s_\vk^g | c^\dagger_{\vk\beta\sigma'} | s_\vk^m \rangle\langle s_\vk^m |c_{\vk\alpha\sigma}| s_\vk^g \rangle}{z - E_{s_\vk^g}(\vk) + E_{s_\vk^m}(\vk)} \right)\,,
\end{equation}
in which $D_\vk$ stands for the ground state degeneracy of the Hamiltonian $H_\vk$. Fig.~\ref{fig:chern-spectrum}(b) in the main text is numerically computed with Eq.~\ref{eqn:greens-function-spec-decom}.

Similarly, the charge-current susceptibility can also be written as a spectral decomposition:
\begin{equation}\label{eqn:chi-0j-spec-decomp}
    \chi_{0j}(\vq, i\Omega) =  \prod_{\vk'}\frac{1}{D_{\vk'}}\sum_{g\in \mathcal{G}}\sum_m \left( \frac{\langle g| \rho_\vq | m\rangle \langle m | J^j_{-\vq} |g\rangle }{i\Omega + E_g - E_m} - \frac{\langle g | J^j_{-\vq} |m\rangle \langle m | \rho_\vq | g \rangle}{i\Omega - E_g + E_m}\right)\,,
\end{equation}
We first study the basic features of the numerator appeared in the spectral decomposition. Both $J^j_{-\vq}$ and $\rho_\vq$ operators contain $c, c^\dagger$ operators from the whole Brillouin zone, thus have to be treated with caution. For generic many-body eigenstates $|g\rangle$ and $|m\rangle$, the matrix elements appeared in the spectral decomposition have the following form:
\begin{align}
    \langle g| \rho_\vq | m\rangle \langle m | J^j_{-\vq} |g\rangle =& \frac{1}{N_L} \sum_{\vk_1, \vk_2}\langle g |\sum_{\alpha\sigma} c^\dagger_{\vk_1 + \frac{\vq}{2}\alpha\sigma}c_{\vk_1 - \frac{\vq}{2}\alpha\sigma}| m\rangle \langle m | \sum_{\alpha'\sigma'\beta\sigma''} v^j_{\alpha'\sigma', \beta,\sigma''}(\vk_2)c^\dagger_{\vk_2-\frac{\vq}{2}\alpha'\sigma'}c_{\vk_2 + \frac{\vq}{2}\beta\sigma''} | g\rangle\,.
\end{align}
Now we analyze the condition of obtaining a non-zero element. The  matrix element $\langle m |c^\dagger_{\vk_2-\frac{\vq}{2}\alpha'\sigma'}c_{\vk_2 + \frac{\vq}{2}\beta\sigma''} | g\rangle$ in the above equation indicates that the state $|m\rangle$ must have {\it one more} electron at $\vk_2 - \frac{\vq}{2}$, and {\it one less} electron at $\vk_2 + \frac{\vq}{2}$ than the state $|g\rangle$. At any other momentum $\vk' \neq \vk \pm \frac{\vq}{2}$, the state $s^m_{\vk'}$ has to be identical to $s^g_{\vk'}$. Otherwise, the matrix element will simply be zero. Due to the same reason, the matrix element $\langle g |c^\dagger_{\vk_1 + \frac{\vq}{2}\alpha\sigma}c_{\vk_1 - \frac{\vq}{2}\alpha\sigma}| m\rangle$ being nonzero implies that the same $|m\rangle$ state has {\it one more} electron at $\vk_1 - \frac{\vq}{2}$, and {\it one less} electron at $\vk_1 + \frac{\vq}{2}$ than the state $|g\rangle$ when the element is non-zero. Therefore, only the terms with $\vk_1 = \vk_2$ will contribute to the susceptibility spectral decomposition. This is a direct consequence of electron number conservation for each $\vk$.

Now we write the states $|g\rangle$ and $|m\rangle$ as the products of eigenstates of $H_\vk$, and the matrix elements will become:
\begin{equation}\label{eqn:mat-element}
    \langle g| \rho_\vq | m\rangle \langle m | J^j_{-\vq} |g\rangle = \frac{1}{N_L} \sum_\vk \prod_{\vk'\neq \vk\pm \frac{\vq}{2}}\delta_{s^g_{\vk'}, s^m_{\vk'}}\langle s^g_{\vk+\frac{\vq}{2}}, s^g_{\vk-\frac{\vq}{2}}|\rho_{\vq, \vk}| s^m_{\vk+\frac{\vq}{2}}, s^m_{\vk - \frac{\vq}{2}}\rangle \langle s^m_{\vk+\frac{\vq}{2}}, s^m_{\vk - \frac{\vq}{2}}|J^j_{-\vq,\vk}| s^g_{\vk+\frac{\vq}{2}}, s^g_{\vk-\frac{\vq}{2}}\rangle\,,
\end{equation}
in which the operators $\rho_{\vq, \vk}$ and $J^j_{\vq, \vk}$ are defined as:
\begin{align}
    \rho_{\vq, \vk} =& \sum_{\alpha\sigma}c^\dagger_{\vk + \frac{\vq}{2}\alpha\sigma}c_{\vk - \frac{\vq}{2}\alpha\sigma}\,,\\
    J^j_{\vq, \vk} =& \sum_{\alpha\sigma,\beta\sigma'}v^j_{\alpha\sigma,\beta\sigma'}(\vk) c^\dagger_{\vk + \frac{\vq}{2}\alpha\sigma} c_{\vk - \frac{\vq}{2}\beta\sigma'}\,.
\end{align}
For each term evolving fermion operators from $\vk \pm \frac{\vq}{2}$, there will also be $\prod_{k'\neq k \pm q/2} D_{k'}$ degenerate ground states who contribute identically to the current correlation function. Therefore, by combining Eq.~\ref{eqn:chi-0j-spec-decomp} and Eq.~\ref{eqn:mat-element}, the susceptibility can eventually be reorganized into the following form:
\begin{align}\label{eqn:chi-0j-hk-spec-decomp}
    &\chi_{0j}(\vq, i\Omega) \nonumber\\
    =& \frac{1}{N_L}\sum_{\vk}\frac{1}{D_{\vk + \frac{q}{2}}D_{\vk - \frac{\vq}{2}}}\sum_{s^g_{\vk+\frac{\vq}{2}}=1}^{D_{\vk+\frac{\vq}{2}}}\sum_{s^g_{\vk-\frac{\vq}{2}}=1}^{D_{\vk-\frac{\vq}{2}}}\sum_{s^m_{\vk\pm\frac{\vq}{2}}=1}^{16}\Bigg{(}\frac{\langle s^g_{\vk+\frac{\vq}{2}},s^g_{\vk-\frac{\vq}{2}} |\rho_{\vq, \vk}| s^m_{\vk + \frac{\vq}{2}}, s^m_{\vk - \frac{\vq}{2}}\rangle \langle s^m_{\vk + \frac{\vq}{2}}, s^m_{\vk - \frac{\vq}{2}} | J^{j}_{-\vq,\vk} | s^g_{\vk+\frac{\vq}{2}},s^g_{\vk-\frac{\vq}{2}} \rangle}{i\Omega + E_{s^g_{\vk+\frac{\vq}{2}}}(\vk + \frac{\vq}{2}) + E_{s^g_{\vk-\frac{\vq}{2}}}(\vk -\frac{\vq}{2}) - E_{s^m_{\vk+\frac{\vq}{2}}}(\vk + \frac{\vq}{2}) -E_{s^m_{\vk-\frac{\vq}{2}}}(\vk - \frac{\vq}{2})}\nonumber\\
    & - \frac{\langle s^g_{\vk+\frac{\vq}{2}},s^g_{\vk-\frac{\vq}{2}} |J^j_{-\vq, \vk}| s^m_{\vk + \frac{\vq}{2}}, s^m_{\vk - \frac{\vq}{2}}\rangle \langle s^m_{\vk + \frac{\vq}{2}}, s^m_{\vk - \frac{\vq}{2}} | \rho_{\vq, \vk} | s^g_{\vk+\frac{\vq}{2}},s^g_{\vk-\frac{\vq}{2}} \rangle}{i\Omega - E_{s^g_{\vk+\frac{\vq}{2}}}(\vk + \frac{\vq}{2}) - E_{s^g_{\vk-\frac{\vq}{2}}}(\vk -\frac{\vq}{2}) + E_{s^m_{\vk+\frac{\vq}{2}}}(\vk + \frac{\vq}{2}) + E_{s^m_{\vk-\frac{\vq}{2}}}(\vk - \frac{\vq}{2})} \Bigg{)}\,.
\end{align}
This susceptibility can also be numerically evaluated using the eigenstates of the many-body Hamiltonian  $H_{\vk}$ for each $\vk$. 
The results shown in Figs.~\ref{fig:chern-spectrum}(c-d) are obtained using Eq.~\ref{eqn:chi-0j-hk-spec-decomp}.

\end{document}